\documentclass[11pt]{article}

\usepackage{ifthen}
\ifthenelse{\isundefined{\compiletoJFstyle}}{
  \newcommand{\compiletoJFstyle}{0} 
}{}
\newcommand{\mmaketitle}{true}
\newcommand{\ttitle}{Misspecified Recovery}
\newcommand{\ttitlefootnote}{We thank Fernando Alvarez, David Backus,  Ravi Bansal, Anmol Bhandari, Peter Carr, Xiaohong Chen,
Ing-Haw Cheng, Mikhail Chernov, Kyle Jurado, Fran\c{c}ois Le Grand, Stavros Panageas, Karthik Sastry, Kenneth Singleton, Johan Walden, Wei Xiong and the anonymous referees for useful comments.}
\newcommand{\aauthorA}{Jaroslav Borovi\v{c}ka}
\newcommand{\aauthorAinstitution}{New York University}
\newcommand{\aauthorAemail}{jaroslav.borovicka@nyu.edu}

\newcommand{\aauthorB}{Lars Peter Hansen}
\newcommand{\aauthorBinstitution}{University of Chicago and NBER}
\newcommand{\aauthorBemail}{lhansen@uchicago.edu \vspace{7mm}}

\newcommand{\aauthorC}{Jos\'{e} A. Scheinkman}
\newcommand{\aauthorCinstitution}{Columbia University, Princeton University and NBER}
\newcommand{\aauthorCemail}{joses@princeton.edu}

\newcommand{\ddate}{October 1, 2015}

\newcommand{\JJELclassification}{C02, C53, C58, D83, G12, G13}
\newcommand{\kkeywords}{long-term risk neutral measure, Perron--Frobenius Theory, recovery, investors' beliefs, stochastic discount factor, martingale}

\newcommand{\JFauthors}{\MakeUppercase{\aauthorA}, \MakeUppercase{\aauthorB} and \MakeUppercase{\aauthorC}}
\newcommand{\JFinstitutions}{Borovi\v{c}ka is with New York University, Hansen is with the University of Chicago and the NBER, and Scheinkman is with Columbia University, Princeton University and the NBER.}
\newcommand{\JFdisclosure}{Borovi\v{c}ka acknowledges financial support from New York University. Hansen's research was supported in part by the Becker--Friedman Institute and by the National Science Foundation, grant 0951576 to the DMUU: Center for Robust Decision Making on Climate and Energy Policy. Scheinkman acknowledges financial support from Columbia University and Princeton University. All authors declare no conflict of interest related to this article.}

\newcommand{\aabstract}{%

Asset prices contain information about the probability distribution of future states and the stochastic discounting of those states as used by investors. To better understand the  challenge in distinguishing investors' beliefs from risk-adjusted discounting, we use Perron--Frobenius Theory to isolate a positive martingale  component of the stochastic discount factor process. This component  recovers a probability measure that absorbs long-term risk adjustments.   When the martingale is not degenerate,  surmising that this recovered probability  captures investors' beliefs distorts  inference about   risk-return tradeoffs. Stochastic discount factors in many structural models of asset prices have empirically relevant martingale components.

}

\newcommand{\pprintcontents}{false}
\newcommand{\separatefrontmatterpages}{true}

\ifthenelse{\equal{\compiletoJFstyle}{1}}{%

\usepackage{sectsty}
\usepackage{amsmath,amssymb,amsthm,amsfonts,enumerate,graphicx}
\usepackage{ifthen,latexsym,syntonly}
\usepackage[normalem]{ulem}
\usepackage{cancel}
\usepackage{color}
\usepackage{pgfplots}
\usepackage{mathtools}
\mathtoolsset{showonlyrefs}

\usepackage{marginnote,subfigure,rotating,fancyvrb}
\usepackage{float}

\usepackage{enumitem} 
\setlist{noitemsep}  
\usepackage[longnamesfirst]{natbib}
\usepackage{datetime}
\usdate
\usepackage{setspace}
\setstretch{1.667}
\setcounter{tocdepth}{2}

\usepackage{indentfirst} 
\usepackage{endnotes}    
\usepackage{jf}          
\usepackage[labelfont=bf,labelsep=period]{caption}   
\captionsetup[table]{labelsep=none}

\newtheorem{condition}{CONDITION}
\newtheorem{corollary}{COROLLARY}
\newtheorem{example}{EXAMPLE}
\newtheorem{proposition}{PROPOSITION}
\newtheorem{problem}{PROBLEM}
\newtheorem{Restriction}{RESTRICTION}

\renewcommand{\footnote}{\endnote}  

\ifthenelse{\isundefined{\theorem}}
{
\newtheorem{theorem}{Theorem}[section]
\newtheorem{remark}[theorem]{Remark}
\newtheorem{assumption}[theorem]{Assumption}
\newtheorem{claim}[theorem]{Claim}
\newtheorem{criterion}[theorem]{Criterion}
\newtheorem{definition}[theorem]{Definition}
\newtheorem{lemma}[theorem]{Lemma}
\newtheorem{result}[theorem]{Result}
\newtheorem{thm}[theorem]{Theorem}
}{}

\newenvironment{exampleenv}[1][Exampleenv]{\noindent\begin{example}}{\makebox[1em][l]{} \hspace{-1em} \hfill $\Box$\end{example}}
\def\proofcaptionfont{\rm\MakeUppercase}

\setlength{\voffset}{0in} \setlength{\topmargin}{0in}
\setlength{\headheight}{0in} \setlength{\headsep}{0in}
\setlength{\textheight}{9in}
\setlength{\hoffset}{0in} \setlength{\oddsidemargin}{0in}
\setlength{\evensidemargin}{0cm}
\setlength{\textwidth}{6.5in}

\newcommand{\be}{\begin{equation}}
\newcommand{\ee}{\end{equation}}

\usepackage[breaklinks]{hyperref} 
\usepackage{ifthen}

\hypersetup{pdfpagemode=UseNone, colorlinks=true, anchorcolor=webbrown, citecolor=webbrown, filecolor= webbrown, linkcolor= webbrown, menucolor= webbrown,
urlcolor= webbrown, citebordercolor=1 0 0, menubordercolor=1 0 0, urlbordercolor=1 0 0, runbordercolor=1 0 0}
\definecolor{webgreen}{rgb}{0,.5,0}
\definecolor{webbrown}{rgb}{.6,0,0}
\definecolor{webblue}{rgb}{0,0,0.8}
\definecolor{webpurple}{rgb}{0.7,0,0.7}

\usepackage{version}
\ifthenelse{\isundefined{\draftnote}}{\newenvironment{draftnote}[1][Draft note]{}{}}{}
\ifthenelse{\isundefined{\includedraftnotes}}{\def\includedraftnotes{false}}{}

\ifthenelse{\equal{\includedraftnotes}{true}}{%
  \renewenvironment{draftnote}[1][Draft note:]%
    {\vspace{2mm}\addtolength{\leftskip}{10mm}\addtolength{\rightskip}{10mm}\begin{small}\noindent\textbf{#1}\ \ }%
    {\hfill\rule{0.5em}{0.5em}\end{small}\par\vspace{2mm}}
}{
  \renewenvironment{draftnote}[1][Draft note:]{}{}
  \excludeversion{draftnote}
}

\usepackage{subfiles}
\ifthenelse{\isundefined{\ssubfile}}{\def\ssubfile{yes}}{}

\ifthenelse{\isundefined{\mmaketitle}}{\def\mmaketitle{false}}{}
\ifthenelse{\isundefined{\ttitle}}{\def\ttitle{}}{}
\ifthenelse{\isundefined{\ttitlefootnote}}{\def\ttitlefootnote{}}{}
\ifthenelse{\isundefined{\aauthor}}{\def\aauthor{}}{}
\ifthenelse{\isundefined{\aauthorfootnote}}{\def\aauthorfootnote{}}{}
\ifthenelse{\isundefined{\aauthorinstitution}}{\def\aauthorinstitution{}}{}
\ifthenelse{\isundefined{\aauthoremail}}{\def\aauthoremail{}}{}
\ifthenelse{\isundefined{\ddate}}{\def\ddate{}}{}
\ifthenelse{\isundefined{\aabstract}}{\def\aabstract{}}{}
\ifthenelse{\isundefined{\pprintcontents}}{\def\pprintcontents{false}}{}
\ifthenelse{\isundefined{\separatefrontmatterpages}}{\def\separatefrontmatterpages{false}}{}

\author{\JFauthors\thanks{\JFinstitutions\ \ttitlefootnote\ \JFdisclosure}}
\title{\Large\bf \ttitle}
\date{}              

\usepackage{abstract}

\newcommand{\mmmaketitle}{%
    \maketitle
    \vspace{-10mm}
    \ifthenelse{\equal{\includedraftnotes}{true}}{%
      \begin{center}{\large Draft with notes. Not for distribution.}\end{center}\par \vspace{5mm}%
    }{}%
    \vfill

    \begin{abstract}
      \onehalfspacing\noindent\aabstract
    \end{abstract}
    
    \bigskip
    
    \bigskip

    \noindent {\it JEL classification:} \JJELclassification
    
    \medskip
    
    {\onehalfspacing
     \noindent {{\it Keywords:} \kkeywords}
     
    }
    
    \vfill
    
    \thispagestyle{empty}\setcounter{page}{0}
    
    \clearpage
    
}

}{%

\usepackage{sectsty}
\usepackage{amsmath,amssymb,amsthm,graphicx}
\usepackage{ifthen,latexsym,syntonly}
\usepackage{setspace}
\usepackage[normalem]{ulem}
\usepackage{cancel}
\usepackage{color}
\usepackage[round,comma,authoryear]{natbib}
\usepackage{pgfplots}
\usepackage{mathtools}
\mathtoolsset{showonlyrefs}
\usepackage{enumitem} 

\bibliographystyle{mynat}

\onehalfspacing

\bibpunct{(}{)}{,}{a}{}{,}  

\ifthenelse{\isundefined{\theorem}}
{
\newtheorem{theorem}{Theorem}[section]

\newtheorem{assumption}[theorem]{Assumption}

\newtheorem{condition}[theorem]{Condition}

\newtheorem{definition}[theorem]{Definition}
\newtheorem{example}[theorem]{Example}

\newtheorem{problem}[theorem]{Problem}
\newtheorem{proposition}[theorem]{Proposition}

}{}

\def\proofcaptionfont{\bf}

\def\mygraphfolder{} 

\setlength{\voffset}{0in} \setlength{\topmargin}{0in}
\setlength{\headheight}{0in} \setlength{\headsep}{0in}
\setlength{\textheight}{9in}
\setlength{\hoffset}{0in} \setlength{\oddsidemargin}{0in}
\setlength{\evensidemargin}{0cm}
\setlength{\textwidth}{6.5in}

\newcommand{\be}{\begin{equation}}
\newcommand{\ee}{\end{equation}}

\usepackage[breaklinks]{hyperref} 
\usepackage{ifthen}

\hypersetup{pdfpagemode=UseNone, colorlinks=true, anchorcolor=webbrown, citecolor=webbrown, filecolor= webbrown, linkcolor= webbrown, menucolor= webbrown,
urlcolor= webbrown, citebordercolor=1 0 0, menubordercolor=1 0 0, urlbordercolor=1 0 0, runbordercolor=1 0 0}
\definecolor{webgreen}{rgb}{0,.5,0}
\definecolor{webbrown}{rgb}{.6,0,0}
\definecolor{webblue}{rgb}{0,0,0.8}
\definecolor{webpurple}{rgb}{0.7,0,0.7}

\usepackage{version}
\ifthenelse{\isundefined{\draftnote}}{\newenvironment{draftnote}[1][Draft note]{}{}}{}
\ifthenelse{\isundefined{\includedraftnotes}}{\def\includedraftnotes{false}}{}

\ifthenelse{\equal{\includedraftnotes}{true}}{%
  \renewenvironment{draftnote}[1][Draft note:]%
    {\vspace{2mm}\addtolength{\leftskip}{10mm}\addtolength{\rightskip}{10mm}\begin{small}\noindent\textbf{#1}\ \ }%
    {\hfill\rule{0.5em}{0.5em}\end{small}\par\vspace{2mm}}
}{
  \renewenvironment{draftnote}[1][Draft note:]{}{}
  \excludeversion{draftnote}
}

\ifthenelse{\isundefined{\mmaketitle}}{\def\mmaketitle{false}}{}
\ifthenelse{\isundefined{\ttitle}}{\def\ttitle{}}{}
\ifthenelse{\isundefined{\ttitlefootnote}}{\def\ttitlefootnote{}}{}
\ifthenelse{\isundefined{\aauthor}}{\def\aauthor{}}{}
\ifthenelse{\isundefined{\aauthorfootnote}}{\def\aauthorfootnote{}}{}
\ifthenelse{\isundefined{\aauthorinstitution}}{\def\aauthorinstitution{}}{}
\ifthenelse{\isundefined{\aauthoremail}}{\def\aauthoremail{}}{}
\ifthenelse{\isundefined{\ddate}}{\def\ddate{}}{}
\ifthenelse{\isundefined{\aabstract}}{\def\aabstract{}}{}
\ifthenelse{\isundefined{\pprintcontents}}{\def\pprintcontents{false}}{}
\ifthenelse{\isundefined{\separatefrontmatterpages}}{\def\separatefrontmatterpages{false}}{}

\ifthenelse{\not\equal{\ttitle}{}}{%
  \ifthenelse{\not\equal{\ttitlefootnote}{}}{%
    \title{{\huge\ttitle}\thanks{\ttitlefootnote}\vspace{5mm}}%
  }{%
    \title{{\huge\ttitle}\vspace{5mm}}%
  }%
}{}%

\ifthenelse{\isundefined{\aauthorA}}{}{%
\ifthenelse{\not\equal{\aauthorA}{}}{%
  \newcommand{\aauthorsA}{{\Large\aauthorA}\vspace{3mm}\\
  {\large\begin{tabular}{c}\aauthorAinstitution \vspace{1mm} \\ \color{webbrown}\aauthorAemail\end{tabular}}}%
  \author{\aauthorsA\vspace{5mm}}%
}{}}%
\ifthenelse{\isundefined{\aauthorB}}{}{%
\ifthenelse{\not\equal{\aauthorB}{}}{%
  \newcommand{\aauthorsAB}{\aauthorsA \and {\Large\aauthorB} \vspace{3mm}\\ {\large\begin{tabular}{c}\aauthorBinstitution \vspace{1mm} \\ \color{webbrown}\aauthorBemail\end{tabular}}}%
  \author{\aauthorsAB\vspace{5mm}}%
}{}}%
\ifthenelse{\isundefined{\aauthorC}}{}{%
\ifthenelse{\not\equal{\aauthorC}{}}{%
  \newcommand{\aauthorsABC}{\aauthorsAB \and {\Large\aauthorC} \vspace{3mm}\\ {\large\begin{tabular}{c}\aauthorCinstitution \vspace{1mm} \\ \color{webbrown}\aauthorCemail\end{tabular}}}%
  \author{\aauthorsABC\vspace{5mm}}%
}{}}%

\date{\vspace{3mm}{\Large\ddate}}

\newcommand{\mmmaketitle}{%
  \ifthenelse{\equal{\mmaketitle}{true}}{%
    \maketitle
    \vspace{-10mm}
    \ifthenelse{\equal{\includedraftnotes}{true}}{%
      \begin{center}{\large Draft with notes. Not for distribution.}\end{center}\par \vspace{5mm}%
    }{}%
    \vfill

    \begin{abstract}
      \aabstract
    \end{abstract}
    \vfill
    \ifthenelse{\equal{\separatefrontmatterpages}{true}}{%
      \thispagestyle{empty}\setcounter{page}{0}
      \newpage%
    }{}%
    \ifthenelse{\equal{\pprintcontents}{true}}{%
      \thispagestyle{empty}
      {\setlength{\baselineskip}{0.8\baselineskip}\tableofcontents}%
      \ifthenelse{\equal{\separatefrontmatterpages}{true}}{%
        \thispagestyle{empty}\setcounter{page}{0}
        \newpage%
      }{}%
    }{}%
  }{}%
}

}

\ifthenelse{\equal{\compiletoJFstyle}{1}}{
  \def\myplotblue{black}
  \def\myplotred{black}
  \def\myplotFillOpacityBlue{0.075}
  \def\myplotFillOpacityRed{0.075}
  \def\myplotFillOpacityDark{0.5}
  \def\myplotFillOpacityLight{0.15}
}{
  \def\myplotblue{blue}
  \def\myplotred{red}
  \def\myplotFillOpacityBlue{0.125}
  \def\myplotFillOpacityRed{0.125}
  \def\myplotFillOpacityDark{0.5}
  \def\myplotFillOpacityLight{0.5}
}
\pgfplotscreateplotcyclelist{mycyclelist}{%
solid\\%
dashed\\%
dashdotted\\%
dotted\\%
dashdotdotted\\%
}%
\pgfplotsset{compat=1.3,width=15cm}%
\def\mytikzscale{0.9}%
\def\myPlotYields{%
\def\myplotFILENAME{\mygraphfolder LRR_example_test_yields.dat}
\def\myplotFILENAMEFILL{\mygraphfolder LRR_example_test_yields_fill.dat}
\pgfplotstableread{\myplotFILENAME}\myplotDATA
\pgfplotstableread{\myplotFILENAMEFILL}\myplotDATAFILL
\begin{tikzpicture}[scale=\mytikzscale]
\begin{axis}[xlabel=maturity (quarters),ylabel=consumption yield to maturity, xmin=0,xmax=100,ymin=0,ymax=0.06,height=0.4\textwidth,width=0.5\textwidth,
grid=major, cycle list name=mycyclelist,legend cell align=left,
y tick label style={/pgf/number format/.cd, scaled y ticks=false, fixed}]

\addplot+[draw=\myplotblue,solid,no markers,line width=0pt,fill=white!0!\myplotblue,draw opacity=1,fill opacity=\myplotFillOpacityDark] table [x index=0, y index=1] {\myplotFILENAMEFILL};

\addplot+[draw=\myplotred,solid,no markers,line width=0pt,fill=white!0!\myplotred,draw opacity=1,fill opacity=\myplotFillOpacityLight] table [x index=0, y index=2] {\myplotFILENAMEFILL};

\addplot+[draw=\myplotblue,solid,no markers,line width=2pt] table [x index=0, y index=1] {\myplotDATA};
\addplot+[draw=\myplotblue,solid,no markers,line width=1pt] table [x index=0, y index=2] {\myplotDATA};
\addplot+[draw=\myplotblue,solid,no markers,line width=1pt] table [x index=0, y index=3] {\myplotDATA};

\addplot+[draw=\myplotred,dashed,no markers,line width=2pt] table [x index=0, y index=4] {\myplotDATA};
\addplot+[draw=\myplotred,dashed,no markers,line width=1pt] table [x index=0, y index=5] {\myplotDATA};
\addplot+[draw=\myplotred,dashed,no markers,line width=1pt] table [x index=0, y index=6] {\myplotDATA};

\end{axis}
\end{tikzpicture}
}%
\def\myPlotBondYields{%
\def\myplotFILENAME{\mygraphfolder LRR_example_test_bond_yields.dat}
\def\myplotFILENAMEFILL{\mygraphfolder LRR_example_test_bond_yields_fill.dat}
\pgfplotstableread{\myplotFILENAME}\myplotDATA
\pgfplotstableread{\myplotFILENAMEFILL}\myplotDATAFILL
\begin{tikzpicture}[scale=\mytikzscale]
\begin{axis}[xlabel=maturity (quarters),ylabel=bond yield to maturity, xmin=0,xmax=100,ymin=0,ymax=0.06,height=0.4\textwidth,width=0.5\textwidth,
grid=major, cycle list name=mycyclelist,legend cell align=left,
y tick label style={/pgf/number format/.cd, scaled y ticks=false, fixed}]

\addplot+[draw=\myplotblue,solid,no markers,line width=0pt,fill=white!0!\myplotblue,draw opacity=1,fill opacity=\myplotFillOpacityDark] table [x index=0, y index=1] {\myplotFILENAMEFILL};

\addplot+[draw=\myplotred,solid,no markers,line width=0pt,fill=white!0!\myplotred,draw opacity=1,fill opacity=\myplotFillOpacityLight] table [x index=0, y index=2] {\myplotFILENAMEFILL};

\addplot+[draw=\myplotblue,solid,no markers,line width=2pt] table [x index=0, y index=1] {\myplotDATA};
\addplot+[draw=\myplotblue,solid,no markers,line width=1pt] table [x index=0, y index=2] {\myplotDATA};
\addplot+[draw=\myplotblue,solid,no markers,line width=1pt] table [x index=0, y index=3] {\myplotDATA};

\addplot+[draw=\myplotred,dashed,no markers,line width=2pt] table [x index=0, y index=4] {\myplotDATA};
\addplot+[draw=\myplotred,dashed,no markers,line width=1pt] table [x index=0, y index=5] {\myplotDATA};
\addplot+[draw=\myplotred,dashed,no markers,line width=1pt] table [x index=0, y index=6] {\myplotDATA};

\end{axis}
\end{tikzpicture}
}%
\def\myPlotXDistUndistorted{%
\def\myplotMaxID{17} 
\begin{tikzpicture}[scale=\mytikzscale]
\begin{axis}[xlabel=conditional volatility $X_2$,ylabel=mean growth rate $X_1$,
xmin=0.5,xmax=1.6,ymin=-0.006,ymax=0.006,
height=0.5\textwidth,width=0.5\textwidth,grid=major, cycle list name=mycyclelist,legend cell align=left,
y tick label style={/pgf/number format/.cd, scaled y ticks=false, fixed, precision=3}
]
\foreach \myplotID in {1,...,\myplotMaxID}{%
\addplot+[draw=\myplotblue,solid,no markers,line width=0pt,fill=white!0!\myplotblue,draw opacity=1,fill opacity=\myplotFillOpacityBlue] table [x index=0, y index=1]
{\mygraphfolder LRR_example_test_X_undist_contours_\myplotID.dat};
}
\end{axis}
\end{tikzpicture}
}%
\def\myPlotXDistDistorted{%
\def\myplotMaxID{13} 
\begin{tikzpicture}[scale=\mytikzscale]
\begin{axis}[xlabel=conditional volatility $X_2$,
xmin=0.5,xmax=1.6,ymin=-0.006,ymax=0.006,
height=0.5\textwidth,width=0.5\textwidth,grid=major, cycle list name=mycyclelist,legend cell align=left,
y tick label style={/pgf/number format/.cd, scaled y ticks=false, fixed, precision=3}
]
\foreach \myplotID in {1,...,\myplotMaxID}{%
\addplot+[draw=\myplotred,solid,no markers,line width=0pt,fill=white!0!\myplotred,draw opacity=1,fill opacity=\myplotFillOpacityRed] table [x index=0, y index=1]
{\mygraphfolder LRR_example_test_X_dist_contours_\myplotID.dat};
}
\addplot+[draw=black,dashed,no markers,line width=1pt] table [x index=0, y index=1]
{\mygraphfolder LRR_example_test_X_dist_contours_rn_1.dat};
\end{axis}
\end{tikzpicture}
}%
\pgfplotscreateplotcyclelist{mycyclelist}{%
solid\\%
dashed\\%
dashdotted\\%
dotted\\%
dashdotdotted\\%
}%
\pgfplotsset{compat=1.3,width=15cm}%
\def\mytikzscale{0.9}%

\begin{document}

\renewcommand{\usepackage}[1]{}

\mmmaketitle


\setcounter{page}{1}



\noindent Asset prices are forward looking and encode information about investors' beliefs.  This leads researchers and policy makers to look at financial market data to gauge the views of the private sector about  the future of the macroeconomy.
It has been known, at least since the path-breaking work of Arrow, that asset prices reflect a combination of investors' risk aversion and the probability distributions used to assess risk. In dynamic models, investors' risk aversion is expressed by stochastic discount factors that include compensations for risk exposures.
In this paper, we ask what can be learned from the Arrow prices about investors' beliefs. Data on asset prices alone are not sufficient to  identify  both the stochastic discount factor and transition probabilities without imposing
additional restrictions.  This additional information could be  time series evidence on the evolution of the Markov state, or it could be information on the market-determined stochastic discount factors.

In a Markovian environment, Perron--Frobenius Theory selects a single transition probability compatible with asset prices. This Perron--Frobenius apparatus has been used in previous research in at least two manners.   First, \cite{hansen_scheinkman:2009} showed that  this tool identifies a probability measure that reflects the long-term implications for risk pricing under rational expectations.   We refer to this  probability as the long-term risk neutral probability since the use of this measure renders the long-term risk-return tradeoffs degenerate.
 \cite{hansen_scheinkman:2009} purposefully distinguish this constructed transition probability from the underlying time series evolution.    The ratio of the recovered to the true probability measure is manifested as a non-trivial martingale component in the stochastic discount factor process.
Second, \cite{ross:2015} applied Perron--Frobenius Theory to identify or to ``recover'' investors' beliefs.  Interestingly, this recovery does not impose rational expectations, thus the resulting Markov evolution could reflect  subjective beliefs of investors and not necessarily the actual time series evolution.

In this paper we  delineate the connection between these seemingly disparate results.  We make clear the special assumptions that are needed to guarantee that the transition probabilities recovered using Perron--Frobenius Theory are equal to the subjective transition probabilities of investors or to the actual probabilities under an assumption of rational expectations. We show that in some often used economic settings --- with permanent shocks to the macroeconomic environment or with investors endowed with recursive preferences --- the recovered probabilities differ from the subjective or actual transition probabilities, and provide a calibrated workhorse asset pricing model that illustrates the magnitude of these differences.


Section~\ref{sec:example}  illustrates the  challenge of identifying the correct probability measure from asset market data in a finite-state space environment.  While the finite-state Markov environment is too constraining for many applications,  the discussion in this section provides an overview of some of the main results in this paper. In particular, we show that:
\begin{itemize}

\item the Perron--Frobenius approach recovers a probability measure that absorbs long-term risk prices;
\item the  density of the Perron--Frobenius probability relative to the physical probability gives rise to a martingale component to the stochastic discount factor process;
\item under rational expectations, the stochastic discount factor process used by \cite{ross:2015} implies that this martingale component is a constant.

\end{itemize}

\noindent To place these results in a substantive context, we provide  prototypical examples of asset pricing models that show that a nontrivial martingale component  arises from (i) permanent shocks to the consumption process or (ii) continuation value adjustment that  appear when investors have recursive utilities.

In subsequent sections we establish these insights in greater generality, a generality rich enough to include many existing structural Markovian models of asset pricing. The framework for this analysis, which allows for continuous state spaces and a richer information structure, is introduced in Section~\ref{sec:setup}.  In Section~\ref{sec:what}, we extend the Perron--Frobenius approach  to this more general setting.   Provided we impose an additional ergodicity condition, this approach identifies a unique probability measure captured by a martingale component to the stochastic discount factor process.

In Section~\ref{sec:LTP}, we show the consequences of using the  probability measure recovered by the use of the Perron--Frobenius Theory when making inferences on the risk-return tradeoff. The recovered probability measure absorbs the martingale component of the original stochastic discount factor and thus the recovered stochastic discount factor is trend stationary.  Since the factors determining long-term risk adjustments are now absorbed in the recovered probability measure, assets  are priced \emph{as if}  long-term risk prices were trivial. This outcome is the reason why we refer to the probability specification revealed by the  Perron--Frobenius approach as the \emph{long-term risk neutral measure}.
Section~\ref{sec:quantitative} illustrates the impact of a martingale component to the stochastic discount factor  in a workhorse asset pricing model that features long-run risk.

Starting in  Section \ref{sec:fip},  we characterize the challenges in identifying subjective beliefs from asset prices.  Initially
 we  pose the \emph{fundamental identification problem}: data on asset prices can only identify the stochastic discount factor up to an arbitrary strictly positive martingale, and thus the probability measure associated with a stochastic discount factor  remains unidentified without imposing additional restrictions or using additional data.
We also extend the analysis of \cite{ross:2015} to this more general setting.    By connecting to the results in Section \ref{sec:what}, we demonstrate in Section~\ref{sec:fip} that the martingale component to the stochastic discount factor process must be identically equal to one  for \cite{ross:2015}'s procedure to reveal the subjective beliefs of investors.  Under these beliefs, the long-term risk-return tradeoff is degenerate. Some might wonder whether the presence of a martingale component could be circumvented in practice by approximating the martingale by a highly persistent stationary process. In Section \ref{sec:additional} we show that when we extend the state vector to address this approximation issue identification of beliefs becomes tenuous.  In Section~\ref{sec:implications}, we provide a unifying discussion of the  empirical approaches that quantify the impact of the martingale components to stochastic discount factors when an econometrician does not use the full array of Arrow prices.  We also suggest other approaches that connect subjective beliefs to the actual time series evolution of the Markov states.    Section~\ref{sec:conclude} concludes.

\section{Illustrating  the identification challenge}\label{sec:example}




There are multiple approaches for extracting probabilities from asset prices.  For instance,  risk neutral probabilities ({e.g., see \cite{ross78} and  \cite{harrisonkreps79}}) and closely related forward measures are frequently used in financial engineering. More recently, Perron--Frobenius Theory has been applied by \cite{bgz}, \cite{hansen_scheinkman:2009} and \cite{ross:2015} to  study asset pricing --- the last two references featuring the construction of an associated probability measure.     \cite{hansen_scheinkman:2009} and \cite{ross:2015} have rather different interpretations of this measure, however.    \cite{ross:2015} identifies this measure with investors' beliefs while \cite{hansen_scheinkman:2009} use it to characterize long-term contributions to risk pricing.  Under this second interpretation,   Perron--Frobenius Theory features an eigenvalue that {\em dominates} valuation
over long investment horizons,  and the resulting probability measure targets the valuation of assets that pay off in the far future as a point of reference.
Following \cite{hansen_scheinkman:2014}, in this section we illustrate the construction of the alternative probability measure using matrices associated with finite-state Markov chains and we explore some simple example economies to understand better the construction of a probability measure based on Perron--Frobenius Theory.

%
%
%
%
%
Let $X$ be a discrete time, $n$-state Markov chain with transition matrix $%
\mathbf{P\,}\dot{=}\,\left[ p_{ij}\right] $.  We suppose that these are the actual
transition probabilities that govern the evolution of the Markov process.
We  identify state $i$ with
a coordinate vector $u^{i}$ with a single one in the $i$-th entry.
The analyst infers the prices of one-period Arrow claims from data. We
represent this input as a matrix $\mathbf{Q}=\left[ q_{ij}\right] $ that
maps a payoff tomorrow specified as a function of tomorrow's state $j$ into
a price in state $i$ today.  Since there are only a finite number of states, the payoff
and price can both be represented as vectors.  In particular, the vector of Arrow prices given the current realization $x$ of the Markov state  is $x'{\mathbf Q}$.  The entries of this vector give the prices of claims payable in each of the possible
states tomorrow.  Any state that cannot be realized tomorrow given the current state $x$ is assigned a price of zero today.

Asset pricing implications are represented conveniently using stochastic discount factors.
Stochastic
discount factor encode adjustments for uncertainty by discounting the next-period state differentially.  Risk premia are larger for states that are more heavily discounted.   In this finite-state Markov environment, we compute
\begin{equation} \label{def:sdiscount}
s_{ij} = {\frac {q_{ij}} {p_{ij}}}
\end{equation}
provided that $p_{ij} > 0$.  The definition of $s_{ij}$ is inconsequential if $p_{ij} = 0$.  Given a matrix
${\mathbf{S\,}}=\,[s_{ij}]$, the stochastic discount factor process has the increment:
\begin{equation} \label{sdincrement}
{\frac {S_{t+1}}{S_t}} = (X_t)' \mathbf{S} X_{t+1}.
\end{equation}
The stochastic discount factor process $S=\{S_{t}:t=0,1,2,...\}$ is initialized at $S_0=1$ and
 accumulates the increments given by \eqref{sdincrement}:
 \begin{equation*}
S_t=\prod_{\tau =1}^{t}\left(X_{\tau -1}\right)^{\prime}{\mathbf{S}
}X_{\tau }.
\end{equation*}
 Observe that $S_t$ depends on the
history of the state from $0$ to $t$.  With this notation, we have two ways to write the period-zero price of  a claim to a vector of state-dependent payoffs $f \cdot X_{t}$ at time~$t$.
\[
{\mathbf Q}^tf \cdot x =  E \left[ S_t \left( f\cdot X_t \right) \mid X_0 = x \right].
\]

Given the matrix ${\mathbf P}$, possibly determined under rational expectations by historical data and stochastic discount factors implied by
an economic model, Arrow prices are given by inverting equation \eqref{def:sdiscount}:
\begin{equation} \label{basicequation}
q_{ij} = s_{ij} p_{ij}.
\end{equation}
A question that we explore is what we can learn about beliefs from Arrow prices. Market sentiments or beliefs are part of the discourse for  both public and private sectors.
We study this question by replacing  the assumption of rational expectations with an assumption
of subjective beliefs.

Unfortunately, there is considerable
flexibility in constructing probabilities from the Arrow prices alone.  Notice that $\mathbf{Q}$ has $n\times n$ entries. $\mathbf{P}$ has $n\times
(n-1)$ free entries because row sums have to add up to one.  In general the stochastic discount factor
introduces $n\times n$ free parameters $s_{ij},\,i,j=1,\dots ,n.$  Since the Arrow
prices are the products given in formula \eqref{basicequation}, there are multiple solutions for
probabilities and stochastic discount factors that are consistent with Arrow prices.\footnote{%
The simple counting requires some qualification when ${\mathbf{Q}}$ has
zeros. For instance, when $q_{ij}=0$, then $p_{ij}=0$ in order to prevent
arbitrage opportunities.}

To depict this flexibility, represent alternative transition probabilities by
\[
{\widetilde p}_{ij} = h_{ij} p_{ij}
\]
where $h_{ij} > 0$ and $\sum_{j=1}^n h_{ij} p_{ij} = 1$ for $i=1,2,...,n$.  Form a matrix ${\mathbf{H}} =
[h_{ij}]$ and a positive process $\{ H_t : t=0,1,... \}$ with increments
\[
{\frac {H_{t+1}}{H_t}} = (X_t)' {\mathbf H} X_{t+1}.
\]
The
restrictions on the entries of ${\mathbf H}$ restrict the increments to satisfy
\[
E\left[ {\frac {H_{t+1}}{H_t}} \mid X_t = x \right] = 1.
\]
Accumulating the increments:
\begin{equation}
H_t= H_0 \prod_{\tau =1}^{t}\left(X_{\tau -1}\right)^{\prime}{\mathbf{H}
}X_{\tau }. \label{eq:HH}
\end{equation}%
The initial distribution of $X_0$ together with the transition matrix $\mathbf P$ define a probability $P$ over realizations of the process X. Because $h_{ij}$ is obtained as a ratio of probabilities, $H$ is a positive
martingale under $P$ for any positive specification of $H_{0}$ as a function
of $X_{0}$.

Using the  positive martingale $H$ to induce a change of measure, we obtain the probability $\widetilde P:$
\[
\widetilde P(X^t= x^t ) = P(X^t= x^t ) H_0 \prod_{\tau =1}^{t}({x_{\tau -1}})^{\prime}{\mathbf{H}}x_{\tau }.
\]
for alternative possible realizations $x^t = (x_0,x_1,...,x_t)$ of $X^t = (X_0,X_1,\dots X_t)$.  In this formula, we presume that
$EH_0 = 1$, and we use $H_0$ in order for $\widetilde P$ to include a change in the initial distribution of $X_0$.  Thus the random variable $H_0$ modifies the distribution of $X_0$ under ${\widetilde P}$ {vis-\`a-vis} $P$, and  $\widetilde{\mathbf P}$ specifies the altered transition probabilities.   Most of our analysis conditions on $X_0$, in which case the choice of $H_0$ is inconsequential and $H_0$ can be set to one for simplicity.
%

For each choice of the restricted matrix ${\mathbf H}$, we may
form the corresponding state-dependent discount factors
${\widetilde s}_{ij} = { {s_{ij}}/{h_{ij}}}$ by applying formula \eqref{def:sdiscount}.   By construction
\begin{equation} \label{recovernew}
q_{ij} = s_{ij} p_{ij} = {\widetilde s}_{ij} {\widetilde p}_{ij}.
\end{equation}
Given flexibility in constructing ${\mathbf H} = \left[ h_{ij}\right]$, we have multiple ways to recover probabilities from
Arrow prices.


We may confront this multiplicity by imposing restrictions on the stochastic discount factors.
As we shall argue, the resulting constructions provide valuable tools for asset pricing even when these probabilities
are not necessarily the same as those used by investors.  In what follows we consider two
alternative restrictions:

\begin{enumerate}[label=(\roman*)]
\item {\em Risk-neutral pricing}:
\begin{equation}
\overline{s}_{i,j} = \overline{q}_i \label{eq:riskneutral}
\end{equation}
for positive numbers $\overline{q}_i, i=1,2,..., n$. This restriction exploits the pricing of one-period discount bonds.


\item {\em Long-term risk pricing}:
\begin{equation}\label{eq:longtermpricing}
{\widehat s}_{ij} = \exp(\eta) {\frac {{ m}_j}{ { m}_i}}
\end{equation}
for positive numbers ${ m}_i, i=1,2,...,n$ and a real number $\eta$ that is typically negative.  The ${ m}_i$'s need only be specified up to a scale factor and the resulting vector can be normalized conveniently.  As we  show below, this restriction helps us characterize long-term pricing implications.

\end{enumerate}

\noindent In both cases we reduce the number of free parameters in the matrix $\mathbf{S}$ from $n^2$ to $n$, making identification of the probabilities possible.  As we show, each approach has an explicit economic interpretation but the matrices of transition probabilities that are recovered do not necessarily coincide with those used by investors or with the actual Markov state dynamics.
%
%
In the first case the difference between the inferred and true probabilities reflects a martingale  that determines the one-period risk adjustments in financial returns. As we show below, in the second case  the difference between the inferred and true probabilities reflects a martingale that determines long-term risk adjustments in pricing stochastically growing cash flows.

\subsection{Risk-neutral probabilities}\label{sec:risk_neutral}

Risk-neutral probabilities are used extensively in the financial engineering literature.   These probabilities are a theoretical construct used to absorb the local or one-period risk adjustments  and are determined by positing a fictitious  ``risk-neutral'' investor. The stochastic discount factor given by \eqref{eq:riskneutral} reflects the fact that all states $j$ tomorrow are discounted equally.
In order to satisfy pricing restrictions \eqref{basicequation}, the risk-neutral transition probabilities must be given by
\begin{equation*}
{\overline p}_{ij}={\frac{q_{ij}}{{\overline{q}}_{i}}}.
\end{equation*}%
Since rows of a probability matrix have to sum up to one, it necessarily follows that ${\overline{q}}_{i}=\sum_{j}^{n}q_{ij}$, which is the price of a one-period discount bond in state $i$.

The risk-neutral probabilities $[{\overline p}_{ij}]$ can always be constructed
and used in conjunction with discount factors $[{\overline s}_{ij}]$. By design the
discount factors are independent of state $j$, reflecting the absence of risk adjustments conditioned on the current state.
In contrast, one-period discount bond prices can still be state-dependent and this dependence is absorbed into the subjective discount factor of the fictitious-risk neutral investor.
While it is
widely recognized that the risk-neutral distribution is distinct from the actual probability distribution, some have argued that the risk-neutral dynamics remain interesting for macroeconomic forecasting precisely because they do embed risk  adjustments.\footnote{Narayana Kocherlakota, President of the Federal Reserve Bank of Minneapolis, during a speech to the Mitsui Financial Symposium in 2012 asks and answers:  ``How can policymakers formulate the needed outlook for marginal net benefits? \ldots\ I argue that policymakers can achieve better outcomes by basing their  outlooks on {\em risk-neutral probabilities} derived from the prices of financial derivatives.''  See \cite{hrr:2014} for a study of public debt using risk-neutral probabilities.}

%

When short-term interest rates are state-dependent, forward measures are sometimes used in valuation. Prices of $t$-period Arrow securities, $q^{[t]}_{ij},$ are the entries of the $t$-th power of the matrix $\mathbf{Q}$.   The $t$-period forward probability measure given the current state $i$ is
\[
\overline{\mathbf{P}}_t= \left[ {\frac {q^{[t]}_{ij}}{\sum_{j=1}^n q^{[t]}_{ij}}} \right].
\]
The denominator used for scaling the Arrow prices is now the price of a $t$-period discount bond.  While the forward measure is of direct interest,
\begin{equation}\label{eq:PtstarPtstar}
\overline{\mathbf{P}}_t  \ne \left(\overline{\mathbf{P}}\right)^t,
\end{equation}
when one-period  bond prices are state dependent. Variation in one-period interest rates contributes to risk adjustment over longer investment horizons, and as a consequence  the construction of  risk-neutral probabilities is horizon-dependent.

%

\subsection{Long-term pricing}\label{subsec:LTP}


We study long-term pricing of  cash flows associated with fixed income securities using Perron--Frobenius Theory.  When there exists a $\lambda >0$ such that the matrix $\sum_{t=0}^{\infty }\lambda ^{t}{\mathbf{Q}}^{t}$  has all entries that are strictly positive, the largest (in absolute value)
eigenvalue  of ${\mathbf{Q}}$ is unique and positive and thus can be written as $\exp ({\widehat \eta} ),$ and has a unique
associated right eigenvector ${\widehat e},$  which has strictly positive entries. Every non-negative eigenvector of $\mathbf Q$ is a scalar multiple of ${\widehat e}.$  We denote the $%
i^{th}$ entry of ${\widehat e}$ as ${\widehat e}_{i}$. Typically, ${\widehat \eta} <0$ to reflect time discounting of future payoffs over long investment horizons.


Recall that we may evaluate \hbox{$t$-period}
claims by applying the matrix ${\mathbf{Q}}$ $t$-times in succession.
From the Perron--Frobenius theory for positive matrices:
\begin{equation*}
\lim_{t\rightarrow \infty }\exp (-{\widehat \eta} t){\mathbf{Q}}^{t}f=(f\cdot {\widehat e}^{\ast }){\widehat e}
\end{equation*}%
where ${\widehat e}^{\ast }$ is the corresponding positive left eigenvector of ${\mathbf{%
Q}}$.   
Applying this formula, the large~$t$ approximation to the rate of discount on an arbitrary security with positive payoff $f\cdot X_t$ in $t$ periods is $-\widehat\eta$.  Similarly, the  one-period holding-period return on this limiting security is: \[
\lim_{t \rightarrow \infty} {\frac {{\mathbf{Q}}^{t-1}f \cdot X_{1}} {{\mathbf{Q}}^{t}f \cdot X_0}} = \exp(- \eta) {\frac {{\widehat e} \cdot X_{1}}
{{\widehat e}\cdot X_0}}.
\]



The eigenvector $\widehat e$ and the associated eigenvalue also provide a way to
construct a probability transition matrix given ${\mathbf{Q}}$. Set
\begin{equation}
\widehat{p}_{ij}:=\exp (-{\widehat \eta} )q_{ij}{\frac{{\widehat e}_{j}}{{\widehat e}_{i}}}.  \label{tilde}
\end{equation}%
Notice that since $\mathbf Q {\widehat e}= \exp ({\widehat \eta} ){\widehat e}$,

\begin{equation*}
\sum_{j=1}^{n}\widehat{p}_{ij}=\exp (-{\widehat \eta} ){\frac{1}{{\widehat e}_{i}}}%
\sum_{j=1}^{n}q_{ij}{\widehat e}_{j}=1.
\end{equation*}%
Thus $\widehat{{\mathbf{P}}}=[\widehat{{p}}_{ij}]$ is a transition
matrix. Moreover,\begin{equation*}
q_{ij}=\exp ({\widehat \eta} ){\frac{{\widehat e}_{i}}{{\widehat e}_{j}}}\widehat{p}_{ij} = \widehat{s}_{ij} \widehat{p}_{ij}.
\end{equation*}
%
Thus we have used the eigenvector $\widehat e$ and the eigenvalue $\widehat \eta$ to  construct a stochastic discount factor that satisfies  \eqref{eq:longtermpricing}  together with a probability measure that satisfies \eqref{basicequation}.
The probability measure constructed in this fashion absorbs the compensations for exposure to long-term components of risk.
Conversely, if one starts with an $\widehat {\mathbf S}$ and $\widehat {\mathbf P}$ that satisfy  \eqref{basicequation} and \eqref{eq:longtermpricing} then it is straightforward to show that the vector with entries $\widetilde e_i= 1 / {m_i}$ is an eigenvector of $\mathbf Q$.\footnote{To see this, notice that $\widetilde{s}_{ij}(1/m_j) = \exp(\eta)(1/m_i)$. The implied probabilities are given by ${\widehat p}_{ij} = {{q_{ij}}/{{\widetilde s}_{{ij}}}}$.
Pre-multiplying by the probabilities $\widehat{p}_{ij}$, summing over $j$, and stacking into the vector form, we obtain $\mathbf{Q}\widetilde e = \exp(\widetilde\eta)\widetilde e$ for a vector $\widetilde e$ with entries $\widetilde e_i = 1/m_i$.}

If we start with the $t$-period Arrow prices in the matrix ${\mathbf Q}^t$  instead of the one-period Arrow prices in the matrix ${\mathbf Q}$, then
\[
{\mathbf Q}^t {\widehat e} = \exp({\widehat \eta} t ) {\widehat e}
\]
for the same vector ${\widehat e}$ and ${\widehat \eta}$.  The implied matrix ${\widehat {\mathbf P}}_t$ constructed from ${\mathbf Q}^t$ satisfies:
\[
{\widehat {\mathbf P}}_t = \left({\widehat {\mathbf P}}\right)^t.
\]
In contrast to the corresponding result \eqref{eq:PtstarPtstar} for risk-neutral probabilities,   Perron--Frobenius Theory  recovers the same $t$-period transition probability regardless whether we use one-period or $t$-period Arrow claims.



\cite{hansen_scheinkman:2009} and \cite{ross:2015} both use this approach to
construct a probability distribution, but they interpret it differently.
\cite{hansen_scheinkman:2009} study multi-period pricing by compounding
stochastic discount factors. 
They use the probability ratios for $\widehat{p}_{ij}$ given by %
\eqref{tilde} and consider the following decomposition:

\begin{equation*}
q_{ij}=\left[ \exp ({\widehat \eta} ){\frac{{\widehat e}_{i}}{{\widehat e}_j}}
{\frac{\widehat{p}_{ij}}{
p_{ij}}}\right] p_{ij}
=\exp ({\widehat \eta} )\left( {\frac{{\widehat e}_{i}}{{\widehat e}_{j}}}\right)
\widehat h_{ij}p_{ij}.
\end{equation*}%
Hence,
\begin{equation}
{s}_{ij}=\exp ({\widehat \eta} )\left( {\frac{{\widehat e}_{i}}{{\widehat e}_{j}}}\right) {\widehat h}_{ij}
\label{eqi:decomp}
\end{equation}%
where  
\[
{\widehat h}_{ij} = \widehat{p}_{ij}/p_{ij}%
\]
provided that $p_{ij} > 0$.  When $p_{ij} = 0$ the construction of ${\widehat h}_{ij}$ is inconsequential.

\cite{hansen_scheinkman:2009} and \cite{hansen:2012} discuss how the decomposition of the
 one-period stochastic discount factor displayed on the right-hand side of \eqref{eqi:decomp}
can be used to study long-term valuation. The third term, which is a ratio
of probabilities, is used as a change of probability measure in their
analysis.  We call this the long-term risk neutral probabilities.  Alternatively, we could follow
 \cite{ross:2015} and use ${\widehat {\mathbf S}} = [ {\widehat s}_{ij} ]$ where
 \[
 {\widehat s}_{ij} = \exp ({\widehat \eta} )\left( {\frac{{\widehat e}_{i}}{{\widehat e}_{j}}}\right)
\]
to construct the stochastic discount factor process and to let ${\mathbf {\widehat P}} = [ {\widehat p}_{ij}]$
denote the subjective beliefs of the investors for the Markov transition.
It is easy to show that ${\widehat h}_{ij}$ cannot be written as ${\widehat h}_{ij}=\exp(\widetilde\eta)\widetilde{e}_i/\widetilde{e}_j$
for some number $\widetilde\eta$ and a vector $\widetilde{e}$ with positive entries, and thus the decomposition in \eqref{eqi:decomp}  is unique.\footnote{For if ${\widehat h}_{ij}=\exp(\widetilde\eta)\widetilde{e}_i/\widetilde{e}_j$ for some number $\widetilde\eta$ and a vector $\widetilde{e}$ with positive entries, there would exist another Perron--Frobenius eigenvector for $\mathbf{Q}$ with entries given by ${\widehat e}_i {\widetilde e}_i$ and an eigenvalue $\exp\left( {\widehat \eta} + {\widetilde \eta} \right)$.  The Perron--Frobenius Theorem guarantees that there is only one eigenvector with strictly positive entries (up to scale)  implying that ${\widehat e}$ must be a vector of constants and ${\widehat \eta} = 0$.}
 In particular we have that:
\[
s_{ij} =\exp ({\widehat \eta} )\left( {\frac{{\widehat e}_{i}}{{\widehat e}_{j}}}\right ) \;\textrm{for some vector}\; {\widehat e} \; \Longleftrightarrow \widehat h_{ij} \equiv 1.
\]

While the asset price data in $\mathbf{Q}$ uniquely determine $(\widehat\eta$, $\widehat e)$  and thus the transition matrix $\mathbf{\widehat P}$, they contain no information about $[\widehat h_{ij}]$ and therefore about $\mathbf{P}$. This highlights the crucial role of restriction~\eqref{eq:longtermpricing}. Additional information or assumptions are needed to separate the right-hand side terms in $\widehat p_{ij} = \widehat h_{ij} p_{ij}$, and imposing $\widehat h_{ij} \equiv 1$ provides such an assumption. Throughout the paper, we study the role of $[\widehat h_{ij}]$ in structural models of asset pricing and ways of identifying it in empirical data.

%

\subsection{Compounding one-period stochastic discounting}

An equivalent statement of equation \eqref{eqi:decomp} is
\[
{\frac{S_{t+1}}{S_{t}}} = \exp(\widehat\eta) \left({\frac {\widehat e \cdot X_t}{\widehat e \cdot X_{t+1}}} \right) \left({\frac{{\widehat H}_{t+1}}{{\widehat H}_{t}}}\right).
\]
Compounding over time and initializing $S_0 = 1$, we obtain
\begin{equation} \label{whatever}
S_t = \exp(\widehat\eta t)  \left({\frac {\widehat e \cdot X_0}{\widehat e \cdot X_t}}\right) \left({\frac {{\widehat H}_t} {{\widehat H}_0}} \right).
\end{equation}
Thus the eigenvalue $\widehat\eta$ contributes an exponential function of $t$
and the eigenvector contributes a function of the Markov state to the stochastic discount factor process.   In addition there is a martingale component $\widehat H$, whose logarithm
has stationary increments. Imposing restriction~\eqref{eq:longtermpricing} on the stochastic discount factor used by investors with subjective beliefs implies that the martingale component under rational expectations is absorbed into the probabilities used by investors.  If investors have rational expectations and \eqref{eq:longtermpricing} is not imposed, the martingale implies a change of measure that absorbs the long-term compensations for exposure to growth rate uncertainty. 

In the next sections we
address these issues under much more generality by allowing for continuous-state Markov processes. As we  will see,  some additional complications
emerge.

\subsection{Examples}\label{subsec:ex}

The behavior of underlying shocks is of considerable interest when constructing stochastic equilibrium models.  There is substantial time series literature on the role of permanent shocks in multivariate analysis and there is a related macroeconomic literature on models with balanced growth behavior,  allowing for stochastic growth.  The martingale components in stochastic discount factors characterize durable components to risk adjustments in valuation over alternative investment horizons. As we will see, one source of these durable components are permanent shocks to the macroeconomic environment.   But valuation models have other sources for this durability, including investors' preferences.
The following examples illustrate that even in this $n$-state Markov-chain context it is possible to obtain a non-trivial martingale component for the stochastic discount factor.

%

 \begin{example} \label{ex:ross}

 Consumption-based asset pricing models assume that the stochastic discount factor process is a representation of investors' preferences over consumption. Suppose that the growth rate of equilibrium consumption is stationary and that investor preferences can be depicted using a power utility function.  For the time being, suppose we impose rational expectations.
 Thus the marginal rate of substitution is
 \[
\exp(-\delta) \left( {\frac {C_{t+1}}{C_t}} \right)^{-\gamma} = \phi\left(X_{t+1}, X_t \right).
 \]
 With this formulation, we may write
 \[
 s_{ij} = \phi\left( X_{t+1} = u_j, X_t = u_i \right).
 \]
 Stochastic growth in consumption as reflected in the entries $s_{ij}$ will induce a martingale component to the stochastic discount factor.  An
 exception occurs when  $C_t = \exp(g_c t) (c \cdot X_t)$ for some vector $c$ with strictly positive entries and a  {known} constant $g_c$  and hence
 \[
 {\frac {C_{t+1}}{C_t}} = \exp(g_c)\left({\frac {c \cdot X_{t+1}}{c \cdot X_t}}\right).
  \]
Here $g_c$ governs the deterministic growth in consumption  and is presumably revealed from time-series data. In this case,
 \[
 s_{ij} = \exp(-\delta-\gamma g_c){\frac {(c_j)^{-\gamma}}{(c_i)^{-\gamma}}}
 \]
 implying that $\widetilde \eta = -(\delta+\gamma g_c)$ and $\widetilde e_j = (c_j)^{\gamma}$.

%
%

Under subjective beliefs and a stochastic discount factor of the form:
\[
{\widehat s}_{ij} = \exp(-\delta-\gamma g_c){\frac {(c_j)^{-\gamma}}{(c_i)^{-\gamma}}},
\]
we may recover subjective probabilities using formula \eqref{tilde}.
This special case is featured in \cite{ross:2015}, but here, except for a deterministic trend, consumption is stationary. Once the consumption process is exposed to  permanent shocks,
the stochastic discount factor inherits a martingale component that reflects this stochastic contribution under the subjective Markov evolution.
 \end{example}

\begin{example}\label{ex:kreps_porteus}

Again let $C_t = \exp(g_c t) (c \cdot X_t)$ be a trend-stationary consumption process where $c$ is an $n \times 1$ vector that represents consumption in individual states of the world.
The (representative) investor is now endowed with recursive preferences of \cite{kreps_porteus:1978} and \cite{epstein_zin:1989}. We consider a special case of unitary elasticity of substitution and initially impose rational expectations.  The continuation value for these preferences satisfies the recursion
\begin{equation}\label{eq:value_recursion1}
V_t = [1-\exp(-\delta)] \log C_t + \frac{\exp(-\delta)}{1-\gamma} \log E_t [\exp((1-\gamma)V_{t+1})],
\end{equation}
where $\gamma$ is a risk aversion coefficient and $\delta$ is a subjective rate of discount. For this example, $V_t = V(t, X_t = u^i) = v_i + g_c t $ where $v_i$ is the continuation value for state $X_t = u^i$ net of a time trend.     Let $v$ be the vector with entry $i$ given by $v_i$ and $  \exp[(1-\gamma) v]$ be the vector with entry $i$ given by $ \exp[(1-\gamma) v_i] $.  The (translated)  continuation values  satisfy the fixed-point equation:
\begin{equation} \label{eq:value_recursion2}
v_i = [1-\exp(-\delta)] \log c_i + {\frac {\exp(-\delta)}{1-\gamma}} \log \left[ \mathbf{P_{i}} \exp[(1-\gamma) v] \right] + \exp(-\delta) g_c
\end{equation}
where $\mathbf{P_{i}}$  is the $i$-th row of the transition matrix $\mathbf{P}$.  This equation gives the current-period continuation
values as a function of the current-period consumption  and the discounted risk-adjusted future continuation values.  We are led to a fixed-point equation because of our interest in an infinite-horizon solution.
Given the solution $v$ of this equation, denote $v^* =  \exp[(1-\gamma) v]$.

The implied stochastic discounting is captured by the following equivalent depictions:
\begin{equation}\label{eq:sij_KP}
s_{ij}  =  \exp[-(\delta + g_c)] \left(\frac{c_i}{c_j}\right)  \left(\frac{v_j^*}{\mathbf{P_{i}}v^*}\right),
\end{equation}
or, compounding over time,
\begin{equation}\label{eq:St_KP}
S_t = \exp[-(\delta + g_c) t] \left(\frac{c \cdot X_0}{c \cdot X_t}\right) \left(\frac{{H}_t^*}{{ H}_0^*}\right)
\end{equation}
where
\begin{equation*}
\frac{{ H}_{t+1}^*}{ {H}_t^*} = \frac{X_{t+1} \cdot v^* }{X_t \cdot \left(\mathbf{P}v^*\right)}.
\end{equation*}
The process $H^*$ is a martingale.  Perron--Frobenius Theory applied to ${\mathbf P}$ implies that ${\mathbf P}v^* = v^*$
if, and only if, $v^*$ has  constant entries.
As long as  the solution $v$ of equation \eqref{eq:value_recursion2} satisfies  $v_i \ne v_j$ for some pair $(i,j)$,  we conclude that ${\mathbf P}v^* \ne v^*$.


For this example
\[
q_{ij} = p_{ij} \exp[-(\delta +g_c)] \left(\frac{c_i}{c_j}\right)  \left(\frac{v_j^*}{\mathbf{P_{i}}v^*}\right).
\]
Solving
\[
\mathbf{Q} \widehat e = \exp(\widehat \eta) \widehat e
\]
for a vector $e$ with positive entries yields
\begin{align*}
\widehat e_j &= c_j, \quad j = 1,\ldots,n \\ 
\widehat \eta &= - (\delta + g_c).
\end{align*}
This Perron--Frobenius solution (\ref{tilde}) recovers the transition matrix  $\mathbf{\widehat{P}}$ given by
\begin{equation*}
\widehat{p}_{ij} = p_{ij} \left(\frac{v_j^*}{\mathbf{P_{i}}v^*}\right).
\end{equation*}
The recovered transition matrix $\mathbf{\widehat{P}}$ absorbs the risk adjustment that arises from fluctuations in the continuation value $v$. In particular, when $\gamma > 1$, transition probabilities $\widehat{p}_{ij}$ are overweighted for low continuation value states $v_j$ next period.  When $\gamma=1$, the two transition matrices coincide because $v^*$ is necessarily constant across states.\footnote{In the limiting case as $\delta\to 0$, the continuation value $v_j$ converges to a constant independent of the state $j$ due to the trend stationarity specification of the consumption process, and the risk adjustment embedded in the potential fluctuations of the continuation values becomes immaterial.}

Consider now the case of subjective beliefs.  Suppose that an analyst mistakenly assumes $\gamma = 1$ even though it is not.  Then the martingale component in the actual stochastic discount factor is absorbed into the probability distribution the analyst attributes to the subjective beliefs.  Alternatively, suppose that $\gamma  > 1$ and the analyst correctly recognizes that beliefs are subjective.  Then
recursion \eqref{eq:value_recursion2} holds with the subjective transition matrix replacing ${\mathbf P}$.
Any attempt to recover probabilities would have to take account of the impact of subjective beliefs on the value function and hence the stochastic discount factor construction.  This impact is in addition
to equation \eqref{recovernew} that links Arrow prices to probabilities and to state-dependent discounting.

\end{example}

\section{General framework}\label{sec:setup}



We now introduce a framework which encompasses a large class of relevant asset pricing models. Consistent intertemporal pricing together with a Markovian property lead us to use a class of stochastic processes called multiplicative functionals. These processes are built from the underlying Markov process in a special way and will be used to model stochastic discount factors.   Alternative structural economic models will imply further restrictions on  stochastic discount factors.

We start with a probability space $\{\Omega ,\mathcal{F},P\}$ and a set of indices $T$ (either the non-negative integers or the
non-negative reals). On this probability space,  there is an $n$-dimensional,  stationary Markov process $X=\{X_{t} :  t \in T \}$ and a $k$-dimensional  process $W$  with increments that are jointly stationary with $X$ and initialized at $W_0 = 0$.
Although we are interested as before in the transition probabilities of the process  $X,$ we use the process $W$ to model the dynamics of $X$ and provide a source for aggregate risks.   The increments to $W$  represent shocks to the economic dynamics and could be independently distributed over time with mean zero. We start with the discrete-time case, postponing the continuous-time case until Section~\ref{sec:framework_ct}. We suppose that
\begin{equation}\label{eq:X_LOM}
X_{t+1} = \phi_x(X_t, \Delta W_{t+1})
\end{equation}
for a known function $\phi_x$ where $\Delta W_{t+1} \doteq W_{t+1} - W_t$. Furthermore we assume:
\begin{assumption}\label{ass:euop}
The process $X$ is ergodic under $P$ and the distribution of $\Delta W_{t+1}$ conditioned on $X_t$ is time-invariant and independent of the past realizations of $\Delta W_s,\; s \le t$ conditioned on $X_t$.
\end{assumption}
\noindent Write ${\mathfrak F} = \left\{ {\mathfrak F}_t  : t \in T \right\}$ for the filtration generated by histories of $W$ and the initial condition $X_0$.

\subsection{Information}


In what follows we assume that $X_t$ is observable at date $t$ but that the shock vector
$ \Delta W_{t+1}$ is not directly observable at date $t+1$.  Many of the results in this paper can be fully understood considering only the case where the shock vector $\Delta W_{t+1}$ can be inferred from $(X_t, X_{t+1}).$ In some examples that we provide later, however,
 the vector $\Delta W_{t+1}$ cannot be inferred from $(X_t, X_{t+1}).$  These are economic models for which there are more sources of uncertainty $\Delta W_{t+1}$ pertinent to investors than there are relevant state variables.  In this case we consider a stationary increment process
 \[
Y_{t+1} - Y_t = \phi_y(X_t, \Delta W_{t+1})
\]
that  together with $(X_{t+1}, X_t)$ reveals $\Delta W_{t+1}$. Thus given
 $X_t$ and knowledge of the functions $\phi_x$ and $\phi_y$, the shock vector  $\Delta W_{t+1}$ can be inferred from $X_{t+1}$ and $Y_{t+1} - Y_t$.
 The observed histories of the joint process $Z\doteq(X,Y)$ thus generate the same filtration ${\mathfrak F}$ as the histories of the shocks $\Delta W$ and the initial condition $X_0.$ Our construction implies that $Z$ is a Markov process with a triangular structure because the  distribution of $(X_{t+1}, Y_{t+1} - Y_t)$ conditioned on ${\mathcal F}_t$ depends only on $X_t$.

\subsection{ Growth, discounting and martingales.}

We introduce a valuable collection of scalar processes $M$ that can be constructed from $Z$.  The evolution of $\log M$ is restricted to have Markov increments of the form:

\begin{condition} \label{r:mrestrict}    $M$ satisfies
\begin{align}
\log M_{t+1} - \log M_t & = \kappa(X_t, \Delta W_{t+1}). \label{discretespecification}
\end{align}
\end{condition}
\noindent Given our invertibility restriction, we may write:
\[
\log M_{t+1} - \log M_t  = \kappa^*(X_t, X_{t+1}, \Delta Y_{t+1})
\]

Processes satisfying Condition~\ref{r:mrestrict}  are restricted versions  of what we call  \emph{multiplicative functionals} of the process $Z.$ (See Appendix \ref{sec:app_multiplicative} for the formal definition of a multiplicative functional.) In what follows we refer to the processes satisfying Condition~\ref{r:mrestrict} simply as multiplicative functionals, because all of the results can be extended to this larger class of processes.

The process $M$ is strictly positive.
For two  such  functionals $M^1$ and $M^2$, the product $M^1M^2$  and the reciprocal $1/M^1$ are also strictly positive multiplicative functionals. Examples of such functionals are the exponential of  linear combinations of the components of $Y.$

In light of Assumption~\ref{ass:euop}, the logarithm of a multiplicative functional $\log M$ has stationary increments, and thus $M$ itself can display geometric growth or decay along stochastic trajectories.  The process $M$ also could be a martingale whose expectation is invariant across alternative forecasting horizons, and in this sense does not grow or decay over time. We use multiplicative  functionals to construct stochastic discount factors, stochastic growth factors  and positive martingales that represent alternative probability measures.

\subsection{An example}


In the following example, we show how multiplicative functionals relate to the Markov chain framework analyzed in Section~\ref{sec:example}. This example consists of a Markov switching model that has state dependence in the conditional  mean and in the exposure to normally distributed shocks. To include this
 richer collection of models, we allow our multiplicative functional to depend on a normally distributed  shock vector $\Delta W$  not fully revealed by the evolution of the state $X.$  Nevertheless,  $X_t$ still serves as the relevant state vector at date $t$.

\begin{example}\label{example:NY}
Let $X$ be a discrete-time, $n$-state Markov chain, and
\[
\Delta W_{t+1} = \begin{bmatrix} X_{t+1} - E\left( X_{t+1} \mid X_t \right) \\ \Delta {\widehat W}_{t+1} \end{bmatrix}
\]
where $\Delta {\widehat W}_{t+1}$ is a $k$-dimensional standard normally distributed random vector that is independent of  $\mathfrak{F}_t$ and $X_{t+1}$.

The first block of the shock vector, $X_{t+1} - E\left( X_{t+1} \mid X_t \right)$, is by construction revealed by the observed realizations of the Markov chain $X$. In addition, we construct the  vector process $Y$ whose $j$-th coordinate evolves as
\[
Y_{j,t+1} - Y_{j,t}  = X_t \cdot \left[ {\bar \mu}_j  +  {\bar \sigma}_j \left(\Delta {\widehat W}_{t+1}\right) \right]
\]
where ${\bar \mu}_j$ is a vector of length $n$, and ${\bar \sigma}_j$ is an $n\times k$ matrix. The matrices  ${\bar \sigma}_j$ are restricted to insure that $\Delta {\widehat W}_{t+1}$ can be computed from $Y_{t+1} - Y_t$ and $X_t$. Thus $Z=(X,Y)$ reveals $W.$
We allow for date $t+1$ Arrow contracts to be written as functions of $Y_{t+1}$ as well as $X_{t+1}$ along with the relevant date $t$ information.
In this environment, we represent the evolution of a multiplicative functional $M$ as:
 \[
\log M_{t+1} - \log M_{t}  = X_t \cdot \left[ {\bar \beta}  +  {\bar \alpha} \left(\Delta {W}_{t+1}\right) \right]
\]
where ${\bar \beta}$ is a vector of length $n$ and ${\bar \alpha}$ is an $n\times (n + k)$ matrix.
\end{example}

%


\subsection{Stochastic discount factors}\label{subsec:sdf}

A \emph{stochastic discount factor process} $S$ is a positive multiplicative
functional with $S_0=1$ and finite first moments (conditioned on $X_{0}$) such that the
date $\tau$ price of any bounded ${\mathfrak F}_t$-measurable claim $\Phi_t$ for $t > \tau$ is:
\begin{equation} \label{eq:pr}
\Pi_{\tau,t}(\Phi_t) \doteq E\left[ {\frac {S_t}{S_\tau} } \Phi_t \mid {\mathfrak F}_\tau \right].
\end{equation}
As a consequence, for a bounded claim $f(X_t)$ that depends only on the current Markov state,
the {time-zero} price is
\begin{equation}\label{eq:Qpricing}
\left[ \mathbb{Q}_{t}f \right](x)\  \doteq E\left[S_{t} f(X_{t})\mid X_{0}=x \right].
\end{equation}
We view $\mathbb{Q}_{t}$ as the pricing operator for payoff horizon $t$.    By construction,
\[
\Pi_{\tau,t}[f(X_t)] =  \left[\mathbb{Q}_{t-\tau} f \right](X_t).
\]
The operator $\mathbb{Q}_t$  is well defined
at least  for bounded functions of the Markov state, but often for a larger class of  functions, depending on the tail behavior of the
stochastic discount factor $S_t$.

The multiplicative property of $S$ allows us to price consistently
at intermediate dates.
In discrete time we can build the $t$-period operator
$\mathbb{Q}_{t}$ by applying the one-period operator $\mathbb{Q}_{1}$ $t$
times in succession and thus
it suffices to study the one-period operator
.\footnote{There is a different stochastic discount factor process that we could use for
much of our analysis.  Let ${\overline {\mathfrak F}}$ denote the (closed) filtration generated  by $X$.  Compute
${\overline S}_t = E\left[ S_t \mid {\overline {\mathfrak F}}_t \right] $.  Then ${\overline S}_t$ is a stochastic discount factor
process pertinent for pricing claims that depend on the history of $X$.  It is also a multiplicative functional constructed from $X$.}

\subsection{Multiplicative martingales and probability measures}


Alternative probability measures equivalent to $P$ are built using
strictly positive martingales. Given an $\mathfrak{F}$-martingale $H$ that
is strictly positive with $E(H_0)=1$,  define a
probability $P^{H}$ such that if $A\in \mathcal{F}_{\tau }$ for some $\tau
\geq 0$,
\begin{equation}
P^{H}(A)=E(1_{A}H_{\tau }).\label{eq:cm}
\end{equation}
The Law of Iterated Expectations guarantees that these definitions are
consistent, that is, if $A\in \mathcal{F}_{\tau }$ and $t>\tau $ then
\begin{equation*}
P^{H}(A)=E(1_{A}H_{t})=E(1_{A}H_{\tau }).
\end{equation*}

Now suppose that $H$ is a multiplicative martingale, a multiplicative functional that is also a martingale with respect to the filtration ${\mathfrak F}$,   modeled as
\[
\log H_{t+1} - \log H_t = h(X_t, \Delta W_{t+1}).
\]
For the martingale restriction to be satisfied, impose
\[
E\left( \exp\left[ h(X_t, \Delta W_{t+1}) \right] \mid X_t= x \right) = 1.
\]
Under the implied change of measure, the probability distribution for $(X_{t+1}, \Delta W_{t+1}) $ conditioned on ${\mathfrak F}_t$ continues to depend only on $X_t.$

While we normalize $S_0 =1$,  we do not do the same for $H_0$.  For some of our subsequent discussion, we
use $H_0$ to alter the initial distribution of $X_0$ in a convenient way.  Thus we allow $H_0$ to depend on $X_0$,
but we restrict it to have expectation equal to unity.

An examination of \eqref{eq:Qpricing} makes
it evident that by using  $S^H=S\frac{H}{H_0}$  as the stochastic discount factor and $P^H$ as the
corresponding probability measure, we will represent the same family of pricing operators $\{ {\mathbb Q}_t : t \ge 0\}$ over bounded functions of $x.$
This flexibility in how we represent pricing extends what we observed in \eqref{recovernew} for the finite-state economies.



\subsection{Continuous-time diffusions}\label{sec:framework_ct}

We impose an analogous structure when $X$ is a continuous-time diffusion.
The process  $W$ is  now an underlying $n$-dimensional Brownian motion and we suppose $X_0$ is independent of $W$  and let $ {\mathfrak F} $ be the (completed) filtration  associated with the Brownian motion augmented to include date-zero information revealed by $Z_0 = (X_0,Y_0)$.     Then  $X$, $Y$ and $\log M$ processes evolve according to:\footnote{While this Brownian information specification abstracts from jumps, these can be included without changing the implications of the analysis, see \cite{hansen_scheinkman:2009}.}
\begin{align}
dX_t & = \mu_x(X_t) dt + \sigma_x(X_t) dW_t \nonumber\\
d Y_t & = \mu_y(X_t) dt + \sigma_y(X_t) dW_t \label{eq:dlogM_ct} \\
d \log M_{t}& =\beta (X_{t})dt+\alpha (X_{t})\cdot dW_{t}. \nonumber
\end{align}%
Notice that the conditional distribution of $(X_{t+\tau}, Y_{t+\tau} - Y_t)$ conditioned on ${\mathfrak F}_t$ depends only on $X_t$ analogous to the assumption that we imposed in the discrete-time specification.  In addition we suppose that
\[
\sigma = \begin{bmatrix} \sigma_x \cr \sigma_y \end{bmatrix}
\]
is nonsingular,
implying that the Brownian motion history is revealed by the $Z:=(X,Y)$ history and ${\mathfrak F}$ is also the filtration associated with the diffusion $Z$.
For the continuous-time specification \eqref{eq:dlogM_ct}, the drift term of a multiplicative martingale $H$ satisfies\footnote{This restriction implies that $H$ is a local martingale and additional restrictions may be required to ensure that $H$ is a  martingale.}
\[
\beta(x) = - {\frac 1 2}  \alpha(x)\cdot \alpha(x).
\]

The definition of a stochastic discount factor and of the family of operators $\mathbb Q_t$ when $t$ is continuous is the direct counterpart to the constructs used in Section \ref{subsec:sdf}  for the discrete-time models.
In continuous time, $\{\mathbb{Q}_{t}:t \in T\}$ forms
what is called a semigroup of operators. The counterpart to a one-period
operator is a \emph{generator} of this semigroup that governs instantaneous
valuation and which acts as a time derivative of ${\mathbb{Q}}_{t}$ at $t=0$.

Under the change of probability induced by $H$, $W_t$ has a drift $\alpha(X_t)$ and is no longer  a martingale.
As in  our discrete-time specification,  under the change of measure,  $Z$ will remain a Markov process and the triangular nature of $Z = (X,Y)$ will be preserved.  Furthermore, we can represent the same operator $\{ {\mathbb Q}_t : t \ge 0 \}$ over bounded functions of $x$ using  $S^H=S\frac{H}{H_0}$  as the stochastic discount factor and $P^H$ as the
corresponding probability measure.

\section{What is recovered}\label{sec:what}

We now review and extend previous results on long-term valuation.  In so doing we  exploit the triangular nature of the Markov process
and feature the state vector $X_t$.  Later we  explore what happens when we extend the state vector to include $Y_t$ in our analysis of long-term pricing.

\subsection{Perron--Frobenius approach to valuation}

Consider a solution to the the following
\emph{Perron--Frobenius} problem:

\begin{problem}[Perron--Frobenius]\label{PFP} Find a scalar ${\widehat \eta}$ and a function ${\widehat e}>0$ such that for every $t \in T,$
\[
\left[{\mathbb Q}_t {\widehat e} \right](x) =  \exp({\widehat \eta} t) {\widehat e}(x).
\]
\end{problem}

\noindent
A solution to this problem  necessarily satisfies the conditional moment restriction:
\begin{equation}\label{larsjunk}
E\left[ S_t {\widehat e}(X_t) \mid {\mathfrak{F}}_\tau \right]  = \exp\left[(t-\tau) {\widehat \eta} \right] S_\tau {\widehat e}(X_\tau)
\end{equation}
for $ t \ge \tau$.    Since ${\widehat e}$ is an eigenfunction, it is only well-defined up to a positive scale factor.  When we make reference to a unique solution to this problem, we mean that ${\widehat e}$ is unique up to  scale.

When the state space is finite as in Section \ref{sec:example}, functions of $x$ can be represented as
vectors in $\mathbb{R}^n,$ and the operator ${\mathbb{Q}}_1$ can be represented as
 a matrix $\mathbf{Q}$. In this case, the existence and uniqueness  of a  solution to Problem~\ref{PFP} is well understood.
Existence and uniqueness are more complicated in the case of general state
spaces. \cite{hansen_scheinkman:2009} present sufficient conditions for the
existence of a solution, but even in examples commonly used in applied work,
multiple (scaled) positive solutions are a possibility. See \cite%
{hansen_scheinkman:2009}, \cite{hansen:2012} and our subsequent discussion for such
examples. If the Perron--Frobenius problem has a solution, we follow
\cite{hansen_scheinkman:2009}, and define a process ${\widehat H}$ that satisfies:
\begin{equation}  \label{junk2}
{\frac {{\widehat H}_t}{{\widehat H}_0}} = \exp( -{\widehat \eta} t) S_t \frac {%
{\widehat e}(X_t)}{{\widehat e}(X_0)}.
\end{equation}
The process ${\widehat H}$ is a positive $\mathfrak{F}$-martingale under the probability measure $P$, since, for $t\ge \tau$,
\begin{equation*}
E\left[{\widehat H}_t \mid \mathfrak{F}_\tau\right]= \frac{\exp(-{\widehat \eta} t)}{{\widehat e}(X_0)}
 E\left[ S_t {\widehat e}(X_t)\mid \mathfrak{F}_\tau \right] {\widehat H}_0%
 = \frac{\exp(-{\widehat \eta} \tau)}{{\widehat e}(X_0)}S_\tau {\widehat e}(X_\tau) {\widehat H}_0 =%
{\widehat H}_\tau,
\end{equation*}
where in the second equality we used equation \eqref{larsjunk}.

The process  ${\widehat H}$ inherits much of the mathematical
structure of the original stochastic discount factor process $S$ and is itself a multiplicative martingale.  For instance, if $S$ has the form given by
Condition~\ref{r:mrestrict}, then:
\begin{align*}
\log {\widehat H}_{t+1} - \log {\widehat H}_t & = \kappa(X_t, \Delta W_{t+1}) + \log
{\widehat e}(X_{t+1}) - \log {\widehat e}(X_t) - {\widehat \eta} \cr & \doteq {\widehat h}(X_t, \Delta W_{t+1})
\end{align*}
where we have used the fact that $X_{t+1} = \phi_x(X_t, \Delta W_{t+1})$.

When we change measures using the martingale ${\widehat H}$,  to be consistent with  the family of pricing operators $\{ {\mathbb Q}_t : t\ge 0\}$ the associated  stochastic discount factor must be:
$$\widehat S_t = S_t \frac{\widehat H_0}{\widehat H_t}= \exp(\widehat \eta) \frac {\widehat e(X_0)}{\widehat e(X_t)}.$$
Under this change of measure, the discounting of a payoff at time $t$ to time 0 is independent of the path of the state between 0 and $t.$

As we
change probability measures, stationarity and ergodicity of $X$ will not necessarily continue to hold.  But checking for this stability under the probability $P^H$ induced by the martingale $H$ will be
featured in  our analysis.  Thus
in establishing a uniqueness  result, we  impose the following condition on the stochastic evolution of $X$ under the probability distribution $P^H$.

\begin{condition}
\label{cond:stationary}
The process $X$ is stationary and ergodic under $P^H.$
\end{condition}

\noindent Stationarity and ergodicity requires the choice of an appropriate  $H_0=h(X_0)$ that induces a stationary distribution under $P^H$ for the Markov process $X$.\footnote{Given ${{H_t}/{H_0}}$, the random variable $H_0 = h(X_0)$ must satisfy the equation:
$$
E\left[ \psi(X_t) \left({\frac {H_t}{H_0}}\right) h(X_0) \right] = E\left[ \psi(X_0)
h(X_0) \right]$$
for any bounded (Borel measurable) $\psi$ and any $t \in T$.}

If $X$ satisfies Condition~\ref{cond:stationary} then it  satisfies a Strong Law of Large Numbers.\footnote{E.g., \cite{breiman}, Corollary 6.23 on page 115.}  In the discrete-time case, if a function $\psi$ has finite expected value, then
\begin{equation*}
\lim_{N \rightarrow \infty}   {\frac 1 {N}} \sum_{t=1}^{N} \psi\left(X_{t}\right) = E^H\psi(X_0)
\end{equation*}
almost surely.
The process $X$ also obeys another version of Law of Large Numbers that considers convergence in means:%
\footnote{Ibid., Corollary 6.25 on page 117.}
\begin{equation*}
\lim_{N \rightarrow \infty} E^H \left[  \left\vert {\frac 1 {N}} \sum_{t=1}^{N} \psi\left(X_{t}\right) -  E^H\psi(X_0) \right\vert \right] = 0,
\end{equation*}
As a consequence of both versions of the Law of Large Numbers, time-series averages of conditional expectations also converge:
 \begin{equation*}
\lim_{N \rightarrow \infty} {\frac 1 {N}} \sum_{t=1}^{N}E^H\left [ \psi\left(X_{t}\right)\mid X_0\right] =   E^H\psi(X_0)
\end{equation*}
almost surely.   Corresponding results hold in continuous time.

We now show that the Perron--Frobenius Problem \ref{PFP} has a unique solution under which $X$ is stationary and ergodic under the probability measure implied by ${\widehat H}$.

\begin{proposition}\label{prop:recoverr}
\label{prop:recover}  There is at most one solution $({\widehat e}, {\widehat \eta})$ to Problem \ref{PFP} such that $X$ is stationary and ergodic under the probability measure $P^{\widehat H}$ induced by  the  multiplicative martingale ${\widehat H}$ given by \eqref{junk2}.
\end{proposition}

\noindent The proof of this theorem is similar to the proof of a related uniqueness result in \cite{hansen_scheinkman:2009} and  is detailed  in Appendix~\ref{sec:app_PF}.\footnote{\cite{hansen_scheinkman:2009} and \cite{hansen:2012} use an implication of the SLLN in their analysis.}  In what follows we use $\widehat P$ and $\widehat E$ instead of the more cumbersome $P^{\widehat H}$ and $E^{\widehat H}$. 


\subsection{An illustration of what is recovered}\label{sec:illustration_recovery}

In the previous discussion, we described two issues arising in the recovery procedure. First, the positive candidate solution for ${\widehat e}(x)$ may not be unique. Our Condition~\ref{cond:stationary} allows us to pick the single  solution that preserves stationarity and ergodicity. Second, even this unique choice may not uncover the true probability distribution if there is a  martingale component in the stochastic discount factor.   The following example shows that in a simplified version of a stochastic volatility model one always recovers an incorrect probability distribution.


\begin{example} \label{ex:sqrootS}
Consider a stochastic discount factor model with state-dependent risk  prices.
\[
 d \log S_t =  {\bar \beta} dt  -  {\frac 1 2 } X_t \left({\bar \alpha} \right)^2 dt +   \sqrt{X_t} {\bar \alpha} dW_t
\]
where ${\bar \beta}  < 0$ and $X$ has the square root dynamics
$$dX_t= -\kappa(X_t- \bar \mu)dt+ \bar \sigma \sqrt{X_t}dW_t.  $$ Guess a solution for a positive eigenfunction:
\[
{\widehat e}(x) = \exp(\upsilon x).
\]
Since $\left\{ \exp( - {\widehat \eta} t) S_t {\widehat e}(X_t) : t \ge 0 \right\}$ is a martingale:
\[
{\bar \beta} - {\frac 1 2} \left({\bar \alpha} \right)^2 x  - \upsilon \kappa x + \upsilon \kappa  {\bar \mu} + {\frac 1 2}  x \left( \upsilon {\bar \sigma} + {\bar \alpha}  \right)^2 - {\widehat \eta} =0.
\]
In particular, the coefficient on $x$ should satisfy
\begin{equation*}
 \upsilon \left[ - \kappa + {\frac 1 2} \upsilon \left( {\bar \sigma} \right)^2 + {\bar \sigma} {\bar \alpha} \right] = 0.
\end{equation*}
There are two solutions: $\upsilon = 0$ and
\begin{equation} \label{hansen1}
\upsilon = {\frac {2 \kappa - 2 {\bar \alpha} {\bar \sigma}}{\left({\bar \sigma} \right)^2}}
\end{equation}

In this  example, the risk neutral dynamics for $X$ corresponds to the  solution $\upsilon = 0$ and the instantaneous risk-free rate is constant and equal to $-{\bar \beta}$.   The resulting $X$ process remains
a square root process, but with $\kappa$ replaced by
\[
\kappa_{n} = \kappa - {\bar \alpha}{\bar  \sigma}.
\]
Although $\kappa$ is positive, $\kappa_n$ could be positive or negative.  If $\kappa_n > 0$, then Condition~\ref{cond:stationary} picks the risk neutral dynamics, which is distinct from the original dynamics for  $X$.  Suppose instead that $\kappa_n < 0$,  which occurs when $\kappa < {\bar \sigma}{\bar \alpha}$.  In this case Condition~\ref{cond:stationary} selects $\upsilon$ given by \eqref{hansen1}, implying that
$\kappa$ is replaced by
\[
\kappa_{pf} = - \kappa + {\bar \sigma}{\bar   \alpha} = -\kappa_n  > 0.
\]
The resulting dynamics are distinct from both the risk neutral dynamics and the original dynamics for  the process $X$.

\end{example}

\noindent This example was designed to keep the algebra simple, but there are straightforward extensions that are described in \cite{hansen:2012}.

Multiplicity of solutions to the Perron--Frobenius problem is prevalent  in models with continuous states. In confronting this multiplicity, \cite{hansen_scheinkman:2009} show that the eigenvalue $\widehat\eta$ that leads to a stochastically stable probability measure $\widehat P$ gives a lower bound to the set of eigenvalues associated with strictly positive eigenfunctions. For a univariate continuous-time Brownian motion setup, \cite{walden:2014} and \cite{park:2014}  construct positive solutions $e$ for every candidate eigenvalue $\eta> \widehat\eta$. However, none of these solution pairs $(e, \eta)$ leads to a probability measure that satisfies Condition~\ref{cond:stationary}.


\section{Long-term pricing}\label{sec:LTP}


Risk neutral probabilities absorb short-term risk adjustments. In contrast, we now show that the probability  measure identified by the application of Perron--Frobenius theory absorbs  risk adjustments over long  horizons.  We call this latter probability measure the long-term risk neutral measure.

Perron--Frobenius Theory features an   eigenvalue ${\widehat \eta}$ and an associated eigenfunction ${\widehat e}$ which determine the limiting behavior of securities with payoffs far in  the future.       We exploit this domination to study long-term risk-return tradeoffs building on the work of \cite{hhl} and \cite{hansen_scheinkman:2009} and long-term holding period returns building on the work of \cite{alvarez_jermann:2005}.  We  show that under the ${\widehat P}$ probability measure, risk-premia on long term cash flows that grow stochastically are zero; but not under the $P$ measure.  We also show that the holding period return on long-term bonds is the increment in the stochastic discount factor under the ${\widehat P}$ measure.

%
%
%

For some of the results in this section, we impose the following refinement of ergodicity.
\begin{condition} \label{stochstable2}  The Markov process $X$ is  aperiodic, irreducible and positive recurrent under the measure ${\widehat P}.$
\end{condition}

\noindent We refer to this condition as {\em stochastic stability}, and it implies that
\[
 \lim_{t\rightarrow \infty} {\widehat E} \left[ f(X_t) \mid X_0 = x \right] = {\widehat E}\left[ f(X_0) \right]
\]
almost surely provided that  ${\widehat E} \left[ f(X_0) \right] < \infty$.\footnote{ For discrete-time models, see \cite{meyntweedie} Theorem 14.0.1 on page 334 for an even stronger conclusion.  We prove the results in this section and appendices only for the discrete-time case. Analogous results for the continuous-time case would use propositions in  \cite{meyntweedieII}.} In this formula, we use the notation ${\widehat E}$ to denote
expectations computed with the probability $\widehat P$ implied by ${\widehat H}$.\footnote{Note that we use $\widehat P$ and $\widehat E$ instead of the more cumbersome $P^{\widehat H}$ and $E^{\widehat H}$.}

%
%


\subsection{Long-term yields} \label{sec:longtermyield}


We first show that the characterization of the eigenvalue ${\widehat \eta}$ in Section \ref{subsec:LTP} extends to this more general framework.    Consider,
\begin{equation*}
\left[{\mathbb{Q}}_t \psi\right](x) = E\left[ S_t \psi(X_t) \mid X_0 = x \right] = \exp(\widehat \eta t) \widehat e(x) {\widehat E} %
\left[ {\frac {\psi(X_t)}{\widehat e(X_t)}} \mid X_0 = x \right]
\end{equation*}
for some positive payoff $\psi(X_t)$ expressed as a function of the Markov state.
For instance, to price pure discount bonds we should set $\psi(x)\equiv 1$.

Consider the implied yield on this investment under the measure ${\widehat P}$:
\begin{equation*}
{\widehat y}_t[\psi(X)](x) \doteq {\frac 1 t} \log {\widehat E} \left[ \psi \left(X_t\right) \mid X_0 = x \right] - {\frac 1 t} \log\left[{\mathbb{Q}}_t \psi\right](x).
\end{equation*}
Taking the limit as $t\to\infty:$
\begin{align}
\label{eq:yield_limit2}
\lim_{t\to\infty} \widehat y_t[\psi(X)](x) &= - \widehat\eta + \lim_{t\to\infty}{\frac 1 t} \log {\widehat E} \left[ \psi \left(X_t\right) \mid X_0 = x \right]  - \lim_{t\to\infty}{\frac1t} \log {\widehat E}
\left[ {\frac {\psi(X_t)}{\widehat e(X_t)}} \mid X_0 = x \right]. \nonumber
\end{align}
This limit shows that $-\widehat \eta$ is the long term yield  maturing in the distant future, provided that the last two terms vanish.
These last two terms vanish under the stochastic stability Condition~\ref{stochstable2} provided that\footnote{Since the logarithms of the conditional expectations are divided by $t$, $ - \widehat\eta$ is the long-term yield under more general circumstances.  See
Appendix~\ref{sec:app_valuation_limits} for details.}
\[
\widehat{E}\left[ \psi \left(X_0\right) \right]  < \infty ,   \qquad \widehat{E} \left[ {\frac {\psi(X_0)}{\widehat e(X_0)} }\right] < \infty.
\]
%


When we use the original measure $P$ instead of ${\widehat P}$, the finite horizon yields will differ.   The limiting yield  still equals $-\widehat \eta$ under the original probability measure, provided that
\[
E \left[\psi(X_t) \right] < \infty.
\]
In summary, stationary cash flow risk does not alter the long-term yield since $-{\widehat \eta}$ is also the yield on a long-term discount bond.  The limiting risk premium is zero under both
probability measures because of the transient nature of the cash-flow risk. Next we introduce payoffs for which stochastic growth implies nonvanishing limiting risk premia.


\subsection{Long-term risk-return tradeoff}\label{sec:long-termrrt}

Consider a positive cash flow process $G$ that grows stochastically over time.   We  model such a cash flow as a multiplicative functional satisfying Condition~\ref{r:mrestrict} with payoff $G_t$ at time $t$.  By design, the growth rate in logarithms  fluctuates randomly depending on the Markov state and the shocks modeled as martingale increments.  Given the multiplicative
nature of $G$, the impact of growth compounds over time.  While we expect the long-term growth rate of $G$ to be positive, the overall exposure to shocks  increases with the
payoff date $t$.    Cash-flow risk is no longer  transient as in Section \ref{sec:longtermyield}.   For convenience we initialize $G_0 =1$.

The yield on the cash flow $G$ under the long-term risk neutral  probability model $(\widehat S, \widehat P)$ is
\begin{equation}\label{eq:distorted_yield}
\widehat y_t[G](x) = \frac1t \log \frac{\widehat E[G_t \mid X_0 = x]}{\widehat E[\widehat S_t G_t \mid X_0 = x]} =  \frac1t \log \frac{E\left[\frac{\widehat H_t}{\widehat H_0} G_t \mid X_0 = x \right]}{ E[S_t G_t \mid X_0 = x]}
\end{equation}
where
\[
\frac{\widehat H_t}{\widehat H_0} = \exp(-\eta t) S_t {\frac {\widehat e(X_t)}{\widehat e(X_0)}} ,
\]
and ${\widehat S}_t = S_t \frac{\widehat H_0}{\widehat H_t}$.
Using this formula for $\widehat H$ in equation \eqref{eq:distorted_yield}, we obtain:
\begin{align*}
\widehat y_t[G](x) &=   \frac1t \log E\left[\frac{\widehat H_t}{\widehat H_0} G_t \mid X_0 = x \right] - \frac1t \log E[S_t G_t \mid X_0 = x]  \\
&=  - \eta + \frac1t \log  E\left[ S_t  {G_t} {\frac {\widehat e(X_t)}{\widehat e(X_0)}}  \mid X_0 = x \right]  - \frac1t \log  E\left[ S_t
 {G_t} \mid X_0 = x \right] \\&= -\eta,
\end{align*}
provided that  additional moment restrictions are imposed. See  Appendix~\ref{sec:app_valuation_limits} for details.
The  limiting yield computed under $\widehat P$ remains the same even after we have introduced stochastic growth in the payoff. In particular, the long-term risk premia on  cash flows  are zero under ${\widehat P}$  even when the cash flows display  stochastic growth.

This conclusion, however, is altered when we compute the yield under the original probability measure.  Now the horizon $t$  yield is:
\[
y_t[G](x) = {\frac 1 t} \log E\left[ G_t \mid X_0 = x \right] - {\frac 1 t}  \log  E \left[ S_t  G_t   \mid X_0 = x \right].
\]
If the martingale components of $S$ and $G$ are correlated over long horizons, the limiting yield of the payoff $G$ under $P$ differs from $-\eta.$

When $S$ and $G$ have non-trivial martingale components, the expected rate of growth of the multiplicative functional $SG$  does \emph{not} typically equal the sum of the  expected rate of growth of $S$ plus the expected rate of growth of $G$. Hence the limiting yield of the payoff $G$ under $P$  differs from $-\eta.$
While the long-term risk premia on stochastically growing cash flows are zero under ${\widehat P}$, these  same long-term risk premia under $P$ are often not degenerate.
By construction, the probability measure associated with Perron--Frobenius Theory makes the long-term risk-return tradeoff vanish.

\subsection{Forward measures and holding-period returns to long-term bonds}\label{sec:holding_period_returns}

Holding-period returns on long-term bonds inform us about the solution to the Perron--Frobenius Problem (Problem~\ref{PFP}) for the  stochastic discount factor process.
The long maturity limit of a holding period return $R_{t,t+1}^\tau$ from period $t$  to period $t+1$  on a bond with maturity $\tau$ is (almost surely)
\begin{equation*}
R_{t,t+1}^\infty = \lim_{\tau \rightarrow \infty} R_{t,t+1}^\tau =
 \lim_{\tau \rightarrow \infty} \frac{[\mathbb{Q}_{\tau-1}{\bf{1}} ](X_{t+1})}{[\mathbb{Q}_{\tau} {\bf 1} ](X_{t})}  =  \exp(-\eta) {\frac {\widehat e(X_{t+1})}{\widehat e(X_t)}}
\end{equation*}
 provided that the stochastic stability Condition~\ref{stochstable2} is imposed and
\[
{\widehat E}\left[{\frac 1 {\widehat e(X_0)}} \right] < \infty.
\]


Since prices of discount bonds at alternative investment horizons are used to construct forward measures, this same  computation  allows us to characterize the limiting forward measure.  As \cite{hansen_scheinkman:2014} argue, the limit of the forward probability measures defined in Section \ref{sec:risk_neutral} above coincides with the measure recovered using  Perron--Frobenius Theory.  To see why,
consider the forward measure at date $t$ for a maturity $\tau.$  We represent this measure using the positive random variable
\[
F_{t,t+\tau} = {\frac {S_{t+\tau}/S_t }{ E \left[S_{t+\tau}/S_t \mid X_t \right]}}
\]
with conditional expectation one given the date-$t$ Markov state $X_t$.  The associated conditional expectations computed using this forward measure
are formed by first multiplying by $F_{t,t+\tau}$ prior to computing the conditional expectations using the $P$ measure.  Thus $F_{t,t+\tau}$ determines the conditional
density of the forward measure with respect to the original measure.
The implied  one-period transition between date $t$ and date $t+1$ is given by
\[
E\left( F_{t,t+\tau} \mid {\mathcal F}_{t+1} \right) = \left( {\frac {S_{t+1}}{S_t} } \right) \left({\frac {E\left[ S_{t+\tau}/S_{t+1} \mid {\mathcal F}_{t+1} \right]}{E \left[S_{t+\tau}/S_t \mid {\mathcal F}_t \right] }}\right)
 =  \left( {\frac {S_{t+1}}{S_t} } \right) {\frac { [\mathbb{Q}_{\tau-1} {\bf 1} ](X_{t+1})}{ [\mathbb{Q}_{\tau} {\bf 1} ](X_{t})}}
\]
which by the Law of Iterated Expectations has expectation equal to unity conditioned on date $t$ information.  Using our previous calculations, taking limits as the investment horizon $\tau$ becomes arbitrarily large, the limiting transition distribution is determined by the random variable:
\begin{equation} \label{eulererror}
\left({\frac {S_{t+1}}{S_t} }\right) R_{t,t+1}^\infty = {\frac {{\widehat H}_{t+1}}{{\widehat H}_t}}
\end{equation}
which reveals the martingale increment in the stochastic discount factor.  This also shows that the limiting one-period transition constructed from the forward measure coincides with the Perron--Frobenius transition  probability.
 \cite{qin_linetsky:2014_1} characterize this limiting  behavior  under more general circumstances without relying on a Markov structure.

When the right-hand side of \eqref{eulererror} is exactly one, the one-period stochastic discount factor is the inverse of the limiting holding-period return.  This link was first noted
by \cite{kazemi:1992}.   More generally, it follows from this formula  that
\begin{equation} \label{logbound}
E\left[ \log R_{t,t+1}^\infty  \mid X_t = x \right]  \le     E\left[  \log S_{t} - \log S_{t+1} \mid X_t = x \right] ,
\end{equation}
since Jensen's Inequality informs us that
\[
 E \left[  \log {\widehat H}_{t+1} - \log { \widehat H}_t \mid X_t = x \right] \le 0.
\]
 \cite{bansallehmann} featured the maximal growth portfolio, that is the portfolio of returns that attains the right-hand side of \eqref{logbound}.
When $ \log {\widehat H}_{t+1} - \log {\widehat H}_t$ is identically zero, $R_{t,t+1}^\infty$ coincides with the return on the maximal growth portfolio.  As \cite{bansal_lehmann:1994}  noted, the \cite{kazemi:1992} result omits permanent components to the stochastic discount factor process.  Nevertheless
the one-period stochastic discount factor can still  be inferred from that maximal growth portfolio provided that an econometrician has a sufficiently rich set of data on returns. Following \cite{alvarez_jermann:2005},  we  use  formula \eqref{eulererror}  in Section~\ref{sec:perm} when we discuss empirical methods and evidence for assessing the magnitude of the martingale component to the stochastic discount factor process.

Now suppose that we change measures and perform the calculations under ${\widehat P}$ using stochastic discount factor:
\[
\widehat S_t = S_t \frac {\widehat H_0} {\widehat H_t}.
\]
In this case
\[
{\widehat E}\left[ \log R_{t,t+1}^\infty  \mid X_t = x \right] =  {\widehat E}\left[  \log {\widehat S}_{t} - \log {\widehat S}_{t+1} \mid X_t = x \right].
\]
To interpret this finding, consider
 any one-period  positive return $R_{t,t+1}$.  Since
\[
{\widehat E} \left[\left({\frac {{\widehat S}_{t+1}}{{\widehat S} _t} }\right) R_{t,t+1} \mid {\mathcal F}_t \right] = 1,
\]
applying Jensen's Inequality,
\begin{equation}\label{eq:Rinfty_max}
{\widehat E} \left[ \log {R}_{t+1} \mid {\mathcal F}_t \right]  \le  E \left[ \log {\widehat S}_{t} - \log {\widehat S}_{t+1}\mid {\mathcal F}_t \right].
\end{equation}
By construction the martingale component of the stochastic discount factor  under the ${\widehat P}$ probability measure is degenerate.  As a consequence,  the inverse of the holding-period return on a long-term bond $R_{t,t+1}^\infty$ coincides with ${ {{\widehat S}_{t+1}}/{{\widehat S} _t} }$ as in \cite{kazemi:1992}'s model.

\section{A quantitative example}\label{sec:quantitative}


We now show that a well-known structural model of asset pricing  proposed by \cite{bansal_yaron:2004}  implies a prominent martingale component.
The
model features growth-rate predictability and
stochastic volatility in the aggregate consumption process. We
utilize a continuous-time Brownian information specification described in
\cite{hhlr} that is calibrated to the consumption dynamics postulated in \cite{bansal_yaron:2004}.

We compare the implications of using the probability measure associated with the Perron--Frobenius extraction with the original probability measure and with the risk neutral measure.
In this example the Perron--Frobenius extraction yields a probability measure that is very similar  to the risk neutral measure and substantially different from the original probability measure.  More generally, our aim in this section is to show that the differences in probability measures could be substantial, rather than giving a definitive conclusion that they are.  The latter conclusion would necessitate a confrontation with direct statistical evidence, an aspect that we discuss in Section~\ref{sec:implications}.

%

Assume the date-$t$ bivariate state vector takes the form $X_{t}=\left( X_{1,t},X_{2,t}\right)^\prime$. In this model,
$X_{1,t}$ represents predictable components in the growth rate of the
multiplicative functional, and $X_{2,t}$ captures the contribution of
stochastic volatility. The dynamics of $X$
specified in \eqref{eq:dlogM_ct} have parameters $\mu \left( x\right) $ and $\sigma \left( x\right) $ given by
\begin{equation}\label{eq:LRR_X_params}
\mu \left( x\right) ={\bar{\mu}}(x-\iota )\qquad \sigma (x)=\sqrt{x_{2}}{%
\bar{\sigma}}
\end{equation}%
where
\begin{equation}
{\bar{\mu}}=\left[
\begin{array}{cc}
\bar{\mu}_{11} & \bar{\mu}_{12} \\
0 & \bar{\mu}_{22}%
\end{array}%
\right] \qquad \bar{\sigma}=\left[
\begin{array}{c}
\bar{\sigma}_{1} \\
\bar{\sigma}_{2}%
\end{array}%
\right].  \label{eq:LRR_X}
\end{equation}%
The parameters $\bar{\sigma}_{1}$ and $\bar{\sigma}%
_{2}$ are $1\times 3$ row vectors. The vector $\iota $ is  the vector of
means of the state variables in a stationary distribution.

All multiplicative functionals $M$ that we consider  satisfy Condition~\ref{r:mrestrict} with parameters $\beta (x) $ and $\alpha (x)$ such that:
\begin{equation}
\beta (x)={\bar{\beta}}_{0}+{\bar{\beta}}_{1}\cdot (x-\iota )\text{\qquad }%
\alpha_x=\sqrt{x_{2}}{\bar{\alpha}}.  \label{eq:LRR_M_params}
\end{equation}%
For instance, the aggregate consumption process $C$ is a multiplicative
functional parameterized by $\left( \beta_{c}(x),\alpha _{c}(x)\right) $.

Appendix~\ref{sec:app_LRR_details} provides details on the
calculations that follow. We use  three uncorrelated shocks in our parameterization. The \emph{direct
consumption} shock is the component of the Brownian motion $W$ that is a
direct innovation to the consumption process $\log C_{t}$. The \emph{growth
rate} shock is the Brownian component that serves as the innovation to the
growth rate $X_{1,t}$, while the \emph{volatility} shock is the innovation
to the volatility process $X_{2,t}$.

We endow the representative investor with the recursive homothetic preferences featured in Example~\ref{ex:kreps_porteus}.  We impose a unitary elasticity of substitution for convenience and thus use the continuous-time counterpart to \eqref{eq:value_recursion1}.  The continuous-time version of these preferences is developed in
\cite{duffie_epstein:1992} and \cite{schroder_skiadas:1999}.

The stochastic discount factor solves
\begin{equation}\label{eq:EZ_SDF}
d\log S_{t}=-\delta dt-d\log C_{t}+d \log H_{t}^*
\end{equation}%
where  $H^*$ is the continuous-time counterpart to the martingale from equation \eqref{eq:St_KP}.\footnote{A rather different motivation for the martingale $H^*$ comes from literature on robustness concerns and asset pricing.  For instance, see \cite{anderson_hansen_sargent:2003}.  In this case $H^*$ is an endogenously determined probability adjustment for potential model misspecification.     Other  models of ambiguity aversion based on max-min utility also induce a martingale component to the stochastic discount factor.}  The $H^*$ component is constructed as follows.  Let $V$ be the forward-looking continuation value process for the homogeneous-of-degree-one utility aggregator, and construct
$ V^{1- \gamma}$ where $\gamma$ is the risk aversion parameter.  The logarithm of the continuation value for the assumed consumption process is an additively separable function of $\log C$ and $X$.   The Brownian increment for the martingale $\bar{S}$ in the stochastic discount factor evolution coincides with the Brownian increment for $V^{1-\gamma}$.
The stochastic discount factor inherits the functional form \eqref{eq:LRR_M_params} with parameters $\left( \beta _{s},\alpha
_{s}\right) $ derived in Appendix~\ref{sec:app_LRR_details}. Since the consumption process $C$ is modeled using a permanent shock, it also contains a martingale component, and thus $H^*$ is not the martingale arising from the Perron--Frobenius problem.

For the Perron--Frobenius probability extraction, we find a solution $(\widehat e,\widehat \eta)$ to the Perron--Frobenius Problem~\ref{PFP} such that
\[
S_t = \exp(\widehat \eta t) \frac{\widehat e(X_0)}{\widehat e(X_t)} \frac{\widehat{H}_t}{\widehat{H}_0} \doteq \widehat S_t \frac{\widehat{H}_t}{\widehat{H}_0}
\]
and $\widehat{H}$ implies a probability measure $\widehat P$ that satisfies Condition~\ref{cond:stationary}.\footnote{The ergodicity and stationarity of $X$ under the recovered measure $\widehat P$, as well as the existence of a solution for the recursive utility stochastic discount factor, can always be checked for given parameters by a direct calculation. For instance, see the calculations in \cite{borovicka_hansen_scheinkman:2014} for details.}

%
%
%
%
%
%
%
%

We show in the appendix that the martingale $\widehat H$ associated with $\widehat P$ takes the form
\begin{equation*}
\frac{d\widehat{H}_{t}}{\widehat{H}_{t}}= \sqrt{X_{2t}}\widehat{\alpha }_{h}\cdot
dW_{t}
\end{equation*}%
where $\widehat{\alpha}_h$ is a vector that depends on the parameters of the model. This implies that we can write the joint dynamics of the state vector $X=(X_1,X_2)^\prime$ as
\begin{eqnarray*}
dX_{1t} &=&\left[ \widehat{\mu }_{11}\left( X_{1t}-\widehat{\iota }%
_{1}\right) +\widehat{\mu }_{12}\left( X_{2t}-\widehat{\iota }%
_{2}\right) \right] dt+\sqrt{X_{2t}}\bar{\sigma}_{1}d\widehat{W}_{t} \\
dX_{2t} &=&\widehat{\mu }_{22}\left( X_{2t}-\widehat{\iota }_{2}\right)
dt+\sqrt{X_{2t}}\bar{\sigma}_{2}d\widehat{W}_{t}
\end{eqnarray*}%
which has the same structure as \eqref{eq:LRR_X_params}--\eqref{eq:LRR_X} with a new set of coefficients $\widehat \mu_{ij}$ derived in the Appendix. The process $\widehat W$ is a Brownian motion under $\widehat P$.

%

\subsection{Forecasts with alternative probability measures}

Structural macro-finance models allow us to construct predictions about the future distribution of macroeconomic quantities and financial cash flows.
Probability measures extracted from asset market data can be used to forecast the future state of the macroeconomy and play a role in the discussion of public policy.  In this section we compare forecasts under the alternative distributions.

\ifthenelse{\equal{\compiletoJFstyle}{1}}{
  \begin{center}
  {\bf [INSERT FIGURE~\ref{fig:X_distribution_latex} AROUND HERE.]}
  \end{center}
}{
  \begin{figure}[!t]
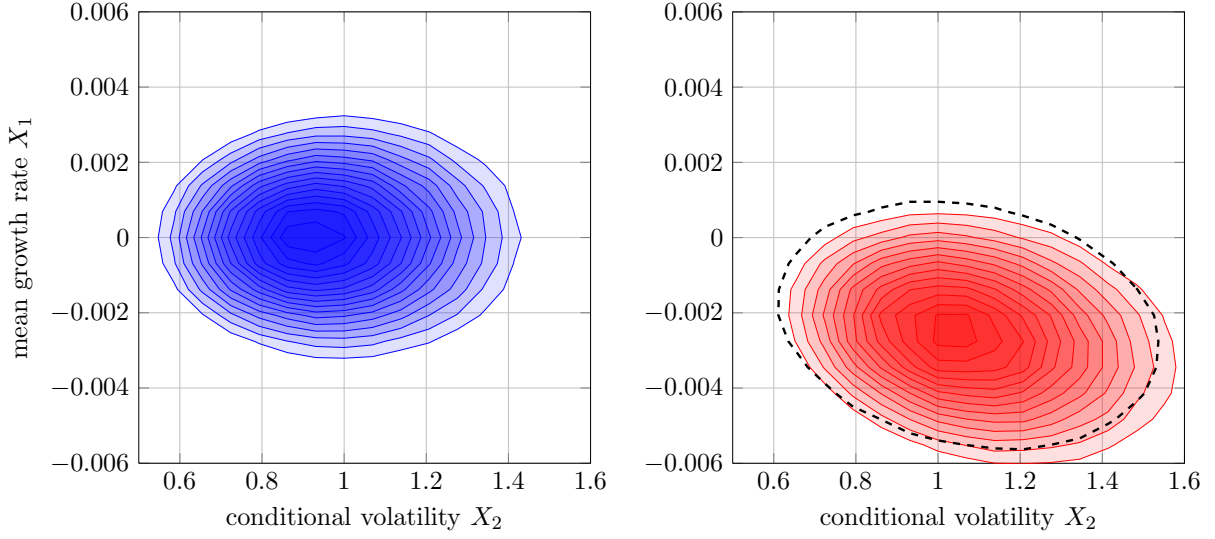

\begin{center}
\myPlotXDistUndistorted\myPlotXDistDistorted
\end{center}
\vspace{-4mm}
\caption{{\bf Stationary densities for the state vector $X=(X_1,X_2)^\prime$ under the correctly specified probability measure $P$ (left panel) and the recovered probability measure $\widehat{P}$ (right panel).} The dashed line in the right panel corresponds to the outermost contour for the distribution under the risk neutral probability measure $\overline{P}$.
The parameterization of the model
is $\bar{\protect\beta}_{c,0}=0.0015$, $\bar{\protect\beta}_{c,1}=1$, $\bar{%
\protect\beta}_{c,2}=0$, $\bar{\protect\mu}_{11}=-0.021$, $\bar{\protect\mu}%
_{12}=\bar{\protect\mu}_{21}=0$, $\bar{\protect\mu}_{22}=-0.013$, $\bar{%
\protect\alpha}_{c}=\left[ 0.0078~~0~~0\right] ^{\prime }$, $\bar{\protect%
\sigma}_{1}=\left[ 0~~0.00034~~0\right] $, $\bar{\protect\sigma}_{2}=\left[
0~~0~~-0.038\right] $, $\protect\iota _{1}=0$, $\protect\iota _{2}=1$, $%
\protect\delta =0.002$, $\protect\gamma =10$. Parameters are calibrated to
monthly frequency.}
\label{fig:X_distribution_latex}
\end{figure}

}


Figure~\ref{fig:X_distribution_latex} plots the joint stationary distribution of the state vector $X$ both under $P$ (left panel) and under $\widehat P$ (right panel). While the distribution in the left panel is the true distribution that is consistent with time series evidence, the distribution in the right panel is the one expected to be observed by a hypothetical investor with beliefs given by $\widehat P$. The distribution under $\widehat P$ exhibits a lower mean growth rate $X_1$ and a higher conditional volatility $X_2$ than the distribution under $P$. Moreover, the adverse states are correlated; low mean growth rate states are more likely to occur jointly with high volatility states. \cite{bidder_smith:2013} document similar distortions in a model with robustness concerns using the martingale $\widehat{S}$ from equation \eqref{eq:EZ_SDF}.


The black dashed line in the right panel of Figure~\ref{fig:X_distribution_latex} gives the outermost contour line for the joint density under the risk neutral dynamics.  The distribution under the risk neutral probability is remarkably similar to the ${\widehat P}$ state probabilities and both are very different from the physical probabilities.

The similarity between the probability measures ${\widehat P}$ and $\overline{P}$ emerges because the martingale component is known to dominate the behavior of the stochastic discount factor.  See \cite{hansen:2012} and \cite{backus_chernov_zin:2014} for evidence to this effect.  Consider the extreme case in which the stochastic discount factor implies that the Perron--Frobenius eigenfunction is constant and the associated martingale implies that under the probability measure ${\widehat P}$ the process $X$ is ergodic.  In this case ${\widehat P} = \overline{P}$, the
short-term interest rate is constant over time and the term structure is flat.  While these term structure implications are not literally true for our parameterized recursive utility model,   the martingale component is sufficiently dominant to imply that risk adjustments embedded ${\widehat P}$ and $\overline{P}$ are very similar.



\subsection{Asset pricing implications}

The probability measures $P$ and $\widehat P$ have substantially different
 implications for yields and holding period returns.
 In Section~\ref{sec:LTP}, we showed that under $\widehat P$, yields on risky cash flows in excess of the riskless benchmark converge to zero as the maturity of these cash flows increases.

\ifthenelse{\equal{\compiletoJFstyle}{1}}{
  \begin{center}
  {\bf [INSERT FIGURE~\ref{fig:LRR_asset_pricing} AROUND HERE.]}
  \end{center}
}{
  \begin{figure}[!t]
\begin{center}
\myPlotYields\myPlotBondYields
\end{center}
\vspace{-4mm}
\ifthenelse{\equal{\compiletoJFstyle}{1}}{
\caption{{\bf Yields under the true and recovered probability measure.} The graphs show the annualized yields on cash flows corresponding to the aggregate consumption process (left panel) and on bonds (right panel) with different maturities. The dark bands with solid lines correspond to the distribution under~$P$, while the light bands with dashed lines to the distribution under~$\widehat P$. The lines represent quartiles of the distribution. The parameterization is as in Figure~\ref{fig:X_distribution_latex}.}
}{
\caption{{\bf Yields under the true and recovered probability measure.} The graphs show the annualized yields on cash flows corresponding to the aggregate consumption process (left panel) and on bonds (right panel) with different maturities. The blue bands with solid lines correspond to the distribution under~$P$, while the red bands with dashed lines to the distribution under~$\widehat P$. The lines represent quartiles of the distribution. The parameterization is as in Figure~\ref{fig:X_distribution_latex}.}
}
\label{fig:LRR_asset_pricing}
\end{figure}


}

The left panel in Figure~\ref{fig:LRR_asset_pricing} plots the yields \eqref{eq:distorted_yield} on a payoff that equals aggregate consumption at $t$ as a function of $t.$  The  solid lines depict the quartiles of the yield distribution ${y}_t[C](x)$ corresponding to the stationary distribution of $X_0 = x$ computed under $P$. The  dashed lines represent the yields $\widehat{y}_t[C](x)$ inferred by an investor who uses the recovered measure $\widehat P$ to compute expected payoffs; but the distribution of these yields is plotted under the correct probability measure $P$ for the current state $X_0 = x$. 

Because the consumption process is negatively correlated with the martingale $\widehat H$, the yields computed under $\widehat P$ are downward biased relative to $P$.
This is necessarily true by construction for long maturities as we show in Section \ref{sec:long-termrrt}, but is also true throughout the term structure.  By construction, the probability measure associated with the  Perron--Frobenius martingale component eliminates  risk adjustments associated with the cash flows from aggregate consumption  at long horizons.  For this example this long-term risk neutral measure accounts for virtually the whole risk premium (in excess of the maturity-matched bond) associated with the cash flows from aggregate consumption at all investment horizons.

\section{Fundamental identification question}\label{sec:fip}

%
%

We now turn to an identification question.  Suppose we observe  Arrow prices for alternative realizations of the Markov state.   Can we recover subjective beliefs?  As we have already observed,
asset prices as depicted by equation \eqref{eq:pr} depend simultaneously on stochastic discount factor processes and on investor beliefs.
A stochastic discount factor process is thus only well defined for a given probability.  If we happen to misspecify investor beliefs, this misspecification can be offset by
altering the stochastic discount factor.  This ability to offset a belief distortion poses a fundamental challenge to the identification
of subjective beliefs.  In this section we formalize this identification problem, and we consider potential restrictions on the stochastic discount factors
that can solve this challenge.




\begin{definition}\label{def:explain_asset_prices}
The pair $(S,P)$ explains asset prices if equation (\ref{eq:pr}) gives the
date zero price of any bounded, ${\mathfrak F}_t$ measurable claim $\Phi_t $ payable at any time $t\in T$.
\end{definition}


%

Consider now a multiplicative martingale $H$ satisfying Condition~\ref{r:mrestrict} and the associated
probability measure $P^{H}$ defined through \eqref{eq:cm}. Similarly let $S$ be a multiplicative functional satisfying Condition~\ref{r:mrestrict} and initialized at
$S_0 = 1$.  We define:
\begin{equation}
S^{H}=S\frac{H_{0}}{H}.\label{eq:SSH}
\end{equation}
The following proposition is immediate:

\begin{proposition}\label{prop:SH_equivalence}
Suppose  $H$ is a  martingale satisfying Condition~\ref{r:mrestrict} with $E \left(H_0 \right) = 1$ and $S$
satisfies  Condition~\ref{r:mrestrict}  with $S_0 = 1$. If the pair $(S, P)$ explains
asset prices then the pair $(S^H, P^H)$ also explains asset prices.  Moreover, $S^H$ satisfies Condition~\ref{r:mrestrict}  and $S^H_0 = 1$.
\end{proposition}

This proposition captures the notion that stochastic discount
factors are only well-defined for a given probability distribution. When we change the probability distribution, we typically must   change the stochastic discount factor
to represent the same  asset prices.  Let $H^1$ and $H^2$ be two positive martingales that satisfy Condition~\ref{r:mrestrict} with
$E \left(H_0^1\right) = E\left(H_0^2\right) = 1$.  Construct the corresponding stochastic discount factors using formula \eqref{eq:SSH}.
Then we cannot distinguish the potential subjective probabilities implied by $H^1$ from those implied by $H^2$
from Arrow prices alone.  There is a pervasive identification problem.
 To achieve identification of investor beliefs, either
we have to restrict the stochastic discount factor process $S$ or we have
to restrict the probability distribution used to represent the valuation
operators $\Pi_{\tau,t}$ for $\tau \le t \in T$.

There are multiple ways we might address this lack of identification. First, we
might impose rational expectations, observe time series data, and let the
Law of Large Numbers for stationary distributions determine the
probabilities. Then observations for a complete set of Arrow securities
allow us to identify $S$.\footnote{See \cite{hansen_richard:1987} for an initial
discussion of the stochastic discount factors and the Law of Large Numbers,
and see \cite{hansensingleton82} for an econometric approach that imposes a parametric structure on the stochastic
discount factor and avoids assuming that the analyst has access to data on the complete set of Arrow securities.}

Alternatively, we may restrict the stochastic discount factor process further.  For instance, risk-neutral pricing
restricts the stochastic discount factor to be predetermined or locally predictable.
Thus for a discrete-time specification:
\[
\log S_{t+1} - \log S_t = \log[\overline q(X_t)]
\]
where $\overline q(X_t)$ is the price of one-period discount bond.  When this restriction is used, typically there is no claim that the resulting probability distribution is the same as that used by
investors.

A different restriction imposes a special structure on $S$:


\begin{condition}
\label{ass:ross}Let $${\widetilde S}_{t}=\exp (-\delta t)\frac{m(X_{t})}{m(X_{0})}$$ for some
positive function $m$ and some real number $\delta.$
\end{condition}

\noindent {\cite{ross:2015} proved an identification result under Condition~\ref{ass:ross} when the dynamics of $X$ are driven by a finite-state Markov chain as in Section~\ref{sec:example}.}  A strengthening of Condition~\ref{ass:ross} is sufficient to guarantee that  Arrow prices identify the stochastic discount factor and a probability distribution associated with that  discount factor  in the more general framework introduced in Section~\ref{sec:setup}.


\begin{proposition} \label{prop:rossagain} Suppose $(S,P)$  explains asset prices.  Let $H$ be a positive multiplicative martingale such that $(S^H,P^H)$ also explains asset prices and $X$ is stationary and ergodic under $P^H$.  If $S^H$ also satisfies Condition~\ref{ass:ross}, then $H$ is uniquely determined.
\end{proposition}

\noindent Thus if  ${\widetilde S}$ satisfies Condition~\ref{ass:ross}, it is the unique $S^H$ identified by this
proposition, and $P^H$ identifies the subjective beliefs ${\widetilde P}$ of investors.  The proof of this result follows directly from our previous analysis.  Let
\[
{\widehat e} = {\frac 1 m}
\]
and $\widehat \eta = -\delta$.   This choice of $({\widehat e}, \widehat \eta)$ solves the Perron--Frobenius Problem \ref{PFP}.  The resulting martingale ${\widehat H}$ leaves the process $X$ as stationary and ergodic.  The conclusion follows directly from Assumption~\ref{ass:euop} and the uniqueness of the solution to  Problem \ref{PFP} established in Proposition \ref{prop:recoverr}.\footnote{In a continuous-time Brownian information setup, alternative conditions on the boundary behavior of the underlying Markov process may also uniquely identify a probability measure. These conditions utilize linkages of the Perron--Frobenius Theorem to the Sturm--Liouville problem in the theory of second-order differential equations. \cite{carr_yu:2012} and \cite{dubynskiy_goldstein:2013} impose conditions on reflecting boundaries, while \cite{walden:2014} analyzes natural boundaries. While  technical conditions like these may deliver a unique solution, they do not resolve the fundamental identification problem once we relax Condition~\ref{ass:ross}.}


Recall from Section~\ref{sec:what} that in general the ${\widehat P}$ probability measure absorbs the long-term risk risk adjustment.  An immediate implication of Proposition \ref{prop:rossagain} is that  Condition~\ref{ass:ross} makes the  long-term risk-return tradeoff degenerate.  Equivalently, the subjective discount factor process has only a degenerate martingale component since the candidate probability measure for depicting subjective beliefs is ${\widetilde P} = {\widehat P}.$    

\section{Additional state vector}\label{sec:additional}

Proposition \ref{prop:recoverr} shows how to identify a martingale associated with the Perron--Frobenius problem. As our examples show, this martingale is, in general, non-trivial.  As a consequence, the ``recovered'' probability measure differs from the subjective probability measure. Perhaps the problem is that we limit  the choice of eigenfunctions too much, by assuming they depend only on the state vector $X_t,$ and not on $Y_t$.  As we show, relaxing this restriction on the eigenfunction could allow for the subjective probability to correspond to one solution to the eigenfunction problem, but we may lose identification even if we impose ergodicity on $X$ and stationary and ergodic increments on $Y$.

We also examine what happens when $Y$ is highly persistent but stationary or stationary around a trend line.  We study this phenomenon by exploring what happens when we approximate a process with stationary increments using highly persistent stationary processes. We argue that, in general, the lack of identification of the limit (stationary increments) process makes it likely that the highly persistent approximations have many \emph{approximate} solutions to the Perron--Frobenius problem.  This phenomenon makes the practical construction of solutions to the Perron--Frobenius problem challenging when it is hard to distinguish a  stationary process from one with stationary increments.\footnote{\cite{walden:2014} documents similar challenges using numerical calculations in both finite- and continuous-state approximations.}
On a more positive note, we show that when we specify a priori a multiplicative process that has the same martingale component as the stochastic discount factor, we may recover subjective beliefs from Arrow prices.

\subsection{Perron--Frobenius revisited}\label{subsec:PFR}

We illustrate the consequences of enlarging the state space in discrete time.  Similar considerations would hold in continuous time. Recall  the joint Markov process:
\begin{align*}
X_{t+1} & = \phi_x(X_t, \Delta W_{t+1}) \cr
Y_{t+1} -  Y_t & = \phi_y(X_t, \Delta W_{t+1}).
\end{align*}
Up until now, we have only allowed for eigenfunctions that depend on $X_t$, but not $Y_t$.  We now entertain eigenfunctions that depend on $(X_t,Y_t)$ when solving the Perron--Frobenius problem.   Thus we now solve:
\[
E\left[ {\frac {S_{t+1}}{S_t}} \varepsilon(X_{t+1},Y_{t+1}) \mid X_t=x, Y_t=y \right] = \exp(\eta) \varepsilon(x,y).
\]
Our previous solution remains a solution to this equation, but there  may be many  others.
To see why, notice that
\[
\exp\left( \zeta \cdot Y \right)
\]
is a multiplicative functional for alternative choices of the vector $\zeta $.
For each choice $\zeta$ solve
\[
E\left[ {\frac {S_{t+1}}{S_t}} \exp\left[ \zeta \cdot g(X_t, \Delta W_{t+1} ) \right] e_\zeta (X_{t+1}) \mid X_t = x \right] =
\exp\left( \eta_\zeta \right) e_\zeta(x) .
\]
As a direct implication of this equation,
\[
E\left[ {\frac {S_{t+1}}{S_t}} \exp\left( \zeta \cdot  Y_{t+1} \right)e_\zeta (X_{t+1}) \mid X_t = x , Y_t = y \right] =
\exp\left( \eta_\zeta \right)  \exp\left(\zeta \cdot  y \right) e_\zeta(x) .
\]
and hence we set
\[
\varepsilon(x,y) = \exp\left( \zeta \cdot y \right) e_\zeta(x).
\]
For each $\zeta$ we could select a solution (should it exist) for which the implied probability measure leaves the state vector process $X$ ergodic and thus $Y$ has stationary and ergodic increments.  But notice that we have constructed a family of solutions indexed by $\zeta$.
While we do not  establish existence, the approaches for doing so remain the same as $S \exp(\zeta \cdot Y)$ is itself a multiplicative functional.  Thus,  augmenting the state vector to include the stationary increment process $Y$ introduces more solutions to the eigenfunction problem, raising the challenge of how to select a particular solution among this family of solutions.

This broader construction of an eigenfunction is of particular interest when
 the following condition is satisfied.


\begin{condition} \label{r:ross3} Let
\[
{\widetilde S}_t = \exp\left( -  {\widetilde \delta} t\right)\exp\left[ {\widetilde \zeta} \cdot (Y_t  - Y_0) \right] \left[\frac {{ \widetilde m}(X_t)}{{\widetilde m}(X_0)} \right]
\]
for some choice of parameters ${ \widetilde \delta}$ and ${\widetilde \zeta}$ and some positive function ${\widetilde m}$ of the state vector $X_t$.
\end{condition}
\noindent For instance when $Y_t= \log C_t,$ Assumption \ref{r:ross3} requires a power utility function perhaps modified by a function of the state vector $X_t,$ that could measure ``habit persistence.''
Given Condition~\ref{r:ross3},
one of the potential Perron--Frobenius eigenfunctions could be used to reveal the subjective probabilities, but we would still be left with the problem in identifying ${\widetilde \zeta}$.  Even with the specificity of Condition~\ref{r:ross3},
there is typically a continuum of solutions to the Perron--Frobenius problem.
Changing $\zeta$ will typically alter the growth rates of the components of $Y$, thus prior restrictions or other information on the subjective rate of growth could be used in conjunction with Arrow prices in a productive way to achieve identification.

\subsection{Stationary approximation}

Highly persistent  stationary processes are hard to distinguish from processes with stationary increments.  While we find it unattractive to exclude a rich class of models that specify stochastic growth, a possible challenge to this view is to argue that model builders should
 focus instead on stationary Markov models that are highly persistent.  In this subsection we suggest that such highly persistent Markov models also present challenges in actual implementation.


We now study a sequence of models that reflect this challenge.  For simplicity suppose that $Y$ is a scalar process.  The multivariate counterpart adds notation but not insight.
Consider a sequence of stationary, ergodic Markov models indexed by $j$:
\begin{align*} \label{embedding}
X_{t+1} & = \phi_x(X_t, \Delta W_{t+1}) \cr
Y_{t+1} - Y_t & = \varphi_y( -\rho_jY_t, X_t, \Delta W_{t+1})
\end{align*}
where $\rho_j > 0 $ converges to zero.  In the limit
\[
Y_{t+1} - Y_t = \varphi_y(0, X_t, \Delta W_{t+1}) = \phi_y(X_t, \Delta W_{t+1}).
\]
For instance, $\varphi_y$ could be affine in its first argument:
\[
\varphi_y( -\rho_jY_t, X_t, \Delta W_{t+1}) = - \rho_j Y_t + \phi_y(X_t, \Delta W_{t+1} ).
\]
We impose the following counterpart to Condition~\ref{ass:ross}.

\begin{condition} \label{r:ross2} Let
\[
{\widetilde S}_{t}=\exp (-{\widetilde \delta} t)\frac {{\widetilde m}(X_{t},Y_t)}{{\widetilde m}(X_{0}, Y_0)}
\]
for some
positive function ${\widetilde m}$ and some real number ${\widetilde \delta}.$
\end{condition}

\noindent For each $j<\infty$, Proposition~\ref{prop:rossagain} guarantees that the Perron--Frobenius problem has a unique solution $\varepsilon(x,y) = { 1 / {{\widetilde m}(x,y)}}$ that preserves the ergodicity of $(X,Y)$.
For the limit problem, Condition~\ref{r:ross2} is not sufficient to guarantee that $S$ is a multiplicative functional satisfying Condition~\ref{r:mrestrict}.  Even after adding Condition~\ref{r:ross3}, as we have already argued,
there is typically a continuum of solutions to the Perron--Frobenius problem for the limiting $\rho_\infty =0$ problem.
This makes approximation challenging as $\rho_j$ becomes close to zero.  The other solutions we described in subsection \ref{subsec:PFR} above  become ``approximate'' solutions when $\rho_j$ is close  to zero.

The formal sense of approximation will matter.  Strictly speaking there is no long-term growth in $Y$ along the sequence even though
the limit process has growth.  Consider the perspective of a researcher who uses standard statistical criteria of approximation that target transition dynamics.
A highly persistent stochastic process will have growth episodes, and thus it is challenging to tell  such a  process from one for which $\Delta Y$ has a positive unconditional mean using standard statistical criteria of approximation that target transition dynamics. To analyze this further would require a more formal discussion of the approximation which is beyond the scope  of this paper.

While this sequence excludes trend growth, the argument could be extended by including a trend line whereby
\[
Y_t^* = Y_t + \nu t
\]
and date $t$ Arrow contracts are written terms of $(Y^*_{t+1}, X_{t+1})$ or $(Y^*_{t+1} - Y_t^*, X_{t+1})$.  Then
\[
Y_{t+1}^* - Y_t^* = Y_{t+1} - Y_t + \nu.
\]
We suppose that our Markov specification of pricing is given in terms of $(Y_{t+1},X_t )$.  If
$\nu$ is known to both the analyst and to investors inside the model, our previous analysis applies.
Alternatively, $\nu$ could be revealed by the time inhomogeneity in the Arrow prices as a function  of the current state.
Either way, the computational challenges we describe previously would  persist.

\subsection{Structured recovery}

We now explore a different
generalization of Condition~\ref{ass:ross} that also delivers a recovery of subjective probabilities.

\begin{condition} The stochastic discount factor process ${\widetilde S}$ satisfies
\[
{\widetilde S}_t = \exp(-\delta t) \left({\frac {Y_t^r} {Y_0^r}} \right) \left[ {\frac {\widetilde m(X_{t})}{\widetilde m(X_0)}} \right]
\]
for some pre-specified multiplicative functional $Y^r$ satisfying Condition~\ref{r:mrestrict}.
\end{condition}
The pre-specified process $Y^r$ captures the long-term risk return tradeoff, and the
stochastic discount factor process ${\widetilde S}$ will have the same martingale component as $Y^r$.
The reciprocal of a positive multiplicative functional is itself a multiplicative functional.  We may restrict
the extended Perron--Frobenius eigenfunction to be of the form
\[
\left( y^r \right)^{-1} e(x).
\]
Once we pre-specify $Y^r$, as in Proposition \ref{prop:rossagain},  $\widetilde m$ and $\delta$ can be inferred from the Arrow prices.
Obtaining uniqueness in these circumstances  amounts to postulating a process that contains the martingale component to consumption.    Given knowledge of $Y^r$, we may again omit the more general vector $y$ from the argument of the eigenfunction in  the Perron--Frobenius calculation. The Arrow prices then reveal the probabilities.

\cite{bansallehmann} and \cite{hansen:2012}  note that in many examples the multiplicative functional
 $Y^r$  could come from a reference model with a direct interpretation.  For instance,  many models of habit persistence, both internal and external have a stochastic discount factor that can be depicted as
\[
{\frac {{\widetilde S}_{t+1}}{{\widetilde S}_t}} = \exp(-\delta) \exp\left[ - \gamma(\log C_{t+1} - \log C_t) \right] {\frac {\widetilde m(X_{t+1})}{\widetilde m(X_t)}}
\]
where $1-\gamma$ is used in the power specification for per period utility and the argument of this utility function depends on a private or social ``habit  stock''.  For this example to be applicable we take\[
 \log Y_t^r =  - \gamma \log C_t
\]
for a known value of $\gamma$.
The function  $m$ of the Markov state captures the impact of the implied non-separability over time in preferences.\footnote{If the habit persistence model is fully parameterized, then given the probabilities one could solve for $m$ as function of these parameters by computing the implied intertemporal marginal rates of substitution.  This would lead to over-identifying restrictions.}  Alternatively,
\cite{hansen_scheinkman:2014} suggest that $m$ can be viewed as a transient misspecification of an underlying asset pricing model-based stochastic discount factor process ${Y}^r$.

\section{Measuring the martingale component}\label{sec:implications}

In this section  we consider methods for extracting evidence from  asset market data about the magnitude of the martingale component in the  stochastic discount factor process. We provide a unifying discussion of the literature and by so doing we add to the existing methods.  This opens the door to new avenues for empirical research.

The existence of the martingale has two interpretations.  Under one interpretation, the assumption of rational expectations allows us to assess the importance of long-term risk adjustments for stochastically growing cash flows.  Under a second interpretation it measures the statistical discrepancy between subjective beliefs and the actual stochastic evolution of the state variables.  For this second interpretation to be valid, we exclude a martingale component in the stochastic discount factor process of the subjective probability model.

\subsection{Quantifying the martingale component} \label{subsec:quantify}

Statistical measures of discrepancy are often constructed using
conveniently chosen convex functions.
Consider functions $\phi_\theta$ defined on the positive real numbers as:
\begin{equation} \label{convex}
\phi_\theta({\sf r}) =   {\frac 1 {\theta (1 +  \theta)}}\left[ \left( {\sf r} \right)^{1 + \theta}  - 1 \right]   \end{equation}
for alternative choices of the parameter $\theta$.  By design $\phi_\theta(1) = 0$ and $\phi''_\theta(1) = 1$.  The function  $\phi_\theta$ remains well defined for $\theta =0$ and $\theta = -1$ by  taking pointwise limits in ${\sf r}$ as $\theta$ approaches these two values.  Thus $\phi_0({\sf r}) =\sf r \log {\sf r}$ and $\phi_{-1}( {\sf r})  = -\log {\sf r}$. The functions $\phi_\theta$ are used to construct discrepancy measures between probability densities as in the work of \cite{cressieread84}.  We are interested in such measures as a way to quantify the martingale component to stochastic discount factors.
Recall that
\[
E\left[{\frac {{\widehat H}_{t+1}}{{\widehat H}_t}} \mid X_t = x \right] =1
\]
and that ${\widehat H}_{t+1}/{\widehat H}_t$ defines a conditional  density of the ${\widehat P}$ distribution relative to the $P$ distribution.  This leads us to apply the discrepancy measures to
${\widehat H}_{t+1}/{\widehat H}_t$.

Since $\phi_\theta$ is strictly convex and $\phi_\theta(1) =0,$ from Jensen's inequality:
\[
E \left[ \phi_\theta \left( {\frac {{\widehat H}_{t+1}}{{\widehat H}_t}}\right) \mid X_t = x \right] \ge 0,
\]
\noindent with equality only when ${\widehat H}_{t+1}/{\widehat H}_t$ is identically one.
There are three special cases that receive particular attention.

\begin{enumerate}[label=(\roman*)]

\item $\theta = 1$ in which case the implied measure of discrepancy is equal to one-half  times the conditional variance of ${\widehat H}_{t+1}/{\widehat H}_t$;

\item $\theta = 0$ in which case  the implied measure of discrepancy is based on conditional relative entropy:
\[
E\left[ \left({\widehat H}_{t+1}/{\widehat H}_t\right) \left( \log {\widehat H}_{t+1} - \log {\widehat H}_t\right) \mid X_t = x \right]
\]
which is the expected log-likelihood under the ${\widehat P}$ probability measure.

\item $\theta = - 1$ in which case the discrepancy measure is:
\[
- E\left[ \log {\widehat H}_{t+1} - \log {\widehat H}_t  \mid X_t = x \right]
\]
which is the negative of the expected log-likelihood under the original probability measure.

\end{enumerate}

We consider two uses of these discrepancy measures.

\subsection{Incomplete asset market data}

Constructing the full range of Arrow prices for alternative states  can be challenging, if not impossible, in practice.
For this reason it is of interest to study empirically what can be learned about stochastic discount factor processes without using an explicit asset pricing model and without using  a complete set of prices of Arrow securities.

We build on the approach initiated by \cite{hansenjagannathan91} aimed at nonparametric characterizations of stochastic discount factors without using a full set of Arrow prices.
While full identification is not possible, data from financial markets remain informative.
We draw on the pedagogically useful characterization of \cite{almeidagarcia13} and \cite{hansennobel:14}, but adapt it to misspecified beliefs along the lines
suggested in \cite{gjt12} and \cite{hansennobel:14}.  In so doing we build on a key insight of \cite{kazemi:1992} and \cite{alvarez_jermann:2005}.

We describe how to compute lower bounds for these discrepancy measures. We are led to the study of lower bounds because we prefer not to compel
 an econometrician to use a full array of Arrow prices.   Let $Y_{t+1}$ be a vector of asset payoffs and $Q_{t}$ the corresponding vector of prices.  Recall the formula for the holding period return on the long term bond
\[
R_{t,t+1}^\infty =    \exp(-\eta) {\frac {e(X_{t+1})}{e(X_t)}},
\]
which implies that
\begin{equation} \label{rewrite}
{\frac {S_{t+1}}{S_t}} = \left({\frac {{\widehat H}_{t+1}}{{\widehat H}_t}}\right) \left({\frac 1 {R_{t,t+1}^\infty}}\right).
\end{equation}
As in \cite{alvarez_jermann:2005} and \cite{bakshi_chabiyo:2012}, suppose that the limiting holding-period return $R_{t,t+1}^\infty$ can be well approximated.
In this case, one could {\em test} directly for the absence  of the martingale component by assessing whether
\[
E\left[  \left({\frac 1 {R_{t,t+1}^\infty}}\right) (Y_{t+1})' \mid X_t = x \right] = \left( Q_t \right)'.
\]
More generally, we express the pricing restrictions as
\[
E\left[  \left({\frac {{\widehat H}_{t+1}}{{\widehat H}_t}}\right) \left({\frac 1 {R_{t,t+1}^\infty}}\right) (Y_{t+1})' \mid X_t = x \right] = \left( Q_t \right)',
\]
where ${\widehat H}$ is now treated as unobservable to an econometrician.
 To bound a discrepancy measure, let a random variable $J_{t+1}$ be a potential specification for the martingale increment:
\[
J_{t+1} =  {\frac {{\widehat H}_{t+1}}{{\widehat H}_t}}.
\] Solve
\[
\lambda_\theta(x) = \inf_{ J_{t+1} > 0} E \left[ \phi_\theta \left(J_{t+1}\right) \mid X_t = x \right]
\]
subject to the linear constraints:
\begin{align*}
E\left[ J_{t+1} \mid X_t =x \right]  - 1 & = 0 \\
E\left[  J_{t+1} \left({\frac 1 {R_{t,t+1}^\infty}}\right) (Y_{t+1})'  \mid X_t = x \right] - \left( Q_t \right)'& = 0.
\end{align*}
A strictly positive $\lambda_\theta(x)$ implies a nontrivial martingale component to the stochastic discount factor.\footnote{A simple computation shows that for the continuous-time diffusion case  the discrepancies  equal  one-half  the local variance of $\log {\widehat H},$  for all values of $\theta$.}

%

To compute $\lambda_\theta$ in practice requires that we estimate conditional distributions.  There is an unconditional counterpart to these calculations obtained by solving:
\begin{equation} \label{simplealternative}
{\bar \lambda_\theta} = \inf_{J_{t+1} > 0} E \phi_\theta(J_{t+1})
\end{equation}
subject to:
\begin{align}
E\left[ J_{t+1}  \right]  - 1 & = 0 \nonumber \\
E\left[  J_{t+1}  \left({\frac 1 {R_{t,t+1}^\infty}}\right) (Y_{t+1})' - (Q_t)' \right] & = 0. \label{priceconstraint}
\end{align}
This bound, while more tractable, is weaker in the sense that ${\bar \lambda_\theta} \le E\lambda_\theta(X_t)$. To
guarantee a solution to optimization  problem \eqref{simplealternative} it is sometimes convenient to  include random variables $J_{t+1}$  that are zero with positive probability.
Since the aim is to produce bounds, this augmentation can be justified for mathematical and computational convenience.    Although this problem optimizes over an infinite-dimensional family of random variables $J_{t+1}$, the dual problem that optimizes over the Lagrange multipliers associated with the pricing constraint \eqref{priceconstraint} is often quite tractable. See \cite{hhl95}  and the literature on implementing generalized empirical likelihood methods for further discussion.\footnote{For the case in which $\theta = 1$,  \cite{hansenjagannathan91} study a mathematically equivalent problem by constructing volatility bounds for stochastic discount factors and deduce quasi-analytical formulas for the solution obtained when ignoring the restriction that  stochastic discount factors  should be nonnegative.  \cite{bakshi_chabiyo:2012} apply the latter methods to obtain  $\theta = 1$ bounds (volatility bounds) for the martingale component of the stochastic discount factor process.  Similarly, \cite{bansallehmann} study bounds on the stochastic discount factor process for the case in which $\theta = -1$ and show the connection with a maximum growth rate portfolio.} \cite{alvarez_jermann:2005} apply these methods to produce the corresponding bounds for the martingale component of the stochastic discount factor process.  While neither paper computes sharp bounds of the type characterized here, both  \cite{alvarez_jermann:2005} and  \cite{bakshi_chabiyo:2012} provide empirical evidence in support of  a substantial martingale component to the stochastic discount factor process using a very similar approach.\footnote{\cite{bakshi_chabiyo:2012} summarize  results from both papers in their Table 1 and   contrast differences in the $\theta =1$ and $\theta = -1$ discrepancy measures.}

\subsection{Arrow prices reconsidered}

In Section~\ref{subsec:PFR}, we provide an extended version of the Perron--Frobenius problem to allow for economies in which there are long-term risk adjustments in valuation.  As a result of this extension, we obtain a parameterized family of solutions even when we observe the full array of Arrow prices.
Recall that a multiplicative martingale is associated with each solution to the Perron--Frobenius problem.  One possibility is to select the parameter for which the implied subjective beliefs are as close as possible to the actual data generating process using one of the measures of statistical discrepancy that we describe in Section~\ref{subsec:quantify}.

There are precursors to such an approach in the literature.  For instance, \cite{stutzer96} uses a $\theta = 0$ relative entropy measure of statistical discrepancy to study  comparisons of risk neutral distributions and in effect Arrow prices to empirical counterparts formed from the actual data generation.
Also, \cite{CCLN:2014} and \cite{christiansen:14} suggest Perron--Frobenius methods for semiparametric identification of stochastic discount factor models under rational expectations, extending the approach of \cite{hansensingleton82}.    \cite{AitsahaliaLo2000} apply a formula from \cite{breedenlitzenberger78} to infer risk neutral densities and actual densities using nonparametric statistical methods for low-dimensional Markov state environments.\footnote{See also \cite{GGR2010} for a recent survey of econometric methods for the study of options prices and
\cite{GGR:2011}  for a discussion of the econometric impact of combining a rich set of options prices for a limited number of dates with time series evidence over many periods.  This latter work also imposes rational expectations.}  Insights from this research may further help in crafting an approach to actual implementation that can be formally justified.

\label{sec:perm}

\section{Conclusion}\label{sec:conclude}

Perron--Frobenius Theory applied to  Arrow prices identifies a martingale component to the market-determined stochastic discount factor process.
This martingale component defines a distorted probability measure that absorbs long-term risk adjustments, in the same spirit as the risk neutral probability measure absorbs one-period risk adjustments. We call this measure the long-term risk neutral measure.

One identifying assumption, featured by \cite{ross:2015}, assumes that this martingale component is identical to one under the subjective beliefs of the investors.  In this case, the probability measure recovered by the Perron--Frobenius Theory coincides with the subjective  probability measure. If, however,  the stochastic discount factor process includes a martingale component, then the use of the Perron--Frobenius eigenvalue and function recovers a long-term risk neutral pricing measure that is distorted by this martingale component.\footnote{Recent working papers by \cite{qin_linetsky:2014_1,qin_linetsky:2014_2} provide additional results in the continuous state space environment, with explicit connections to  previous results in \cite{hansen_scheinkman:2009} and \cite{hansen_scheinkman:2014}.}
By expanding the space of functions used in seeking a solution to the Perron--Frobenius problem, we avoid assuming that long-term risk return tradeoffs are degenerate under subjective beliefs, but we inherit an identification problem.  There is typically a whole family of solutions with little guidance as to which solution should be selected by an analyst.

Many structural models of asset pricing that are motivated by empirical evidence have non-trivial martingale components in the stochastic discount factors.  These martingales characterize  what probability is actually recovered by the application of Perron--Frobenius Theory. We illustrate this outcome in  one  example  with long-run risk components to the macroeconomy and  with investors that have non-separable recursive preferences.
We also provide a unifying discussion of the empirical literature that derives non-parametric bounds for the magnitude of the martingale component and that finds a quantitatively large role for this martingale component for valuation. Finally, we suggest ways  to further expand the set of testable implications in this literature and to combine information about subjective beliefs from Arrow prices with the observed time series evolution.

In our previous work we showed how Perron--Frobenius Theory helps us understand risk-return tradeoffs.
The probability measure identified by Perron--Frobenius Theory absorbs the long-term risk adjustments.  Its naive use can distort the risk-return tradeoff in unintended ways.    One might argue, however, that the dynamics under the Perron--Frobenius probability measure are of interest precisely because this measure adjusts for the long-term riskiness of the macroeconomy.  While we see value in using this probability measure prospectively, our analysis makes clear that the resulting forecasts are slanted in a particular but substantively interesting way.

Finally,  long-term valuation is only a component to a more systematic study of pricing  implications over alternative investment horizons.  Recent  work by \cite{bhhs11} and \cite{borovicka_hansen_scheinkman:2014} deduces methods that extend impulse response functions to characterize the pricing of exposures to  shocks to stochastically growing cash  flows  over alternative investment horizons.

\clearpage
\appendix
\ifthenelse{\equal{\compiletoJFstyle}{1}}{}{%
  \section*{Appendix}


  \begin{small}
  \sectionfont{\large}
  \subsectionfont{\normalsize}
}

\section{Multiplicative functional}\label{sec:app_multiplicative}


The construct of a multiplicative functional is used elsewhere in the probability literature and in \cite{hansen_scheinkman:2009}.
Formally  a {\em multiplicative functional} is a process $M$ that is adapted to $\mathfrak{%
F}$, initialized at $M_{0}=1$ and, with a slight abuse of notation:
\begin{equation}
{M_{t}(Y)}=M_{\tau }(Y)M_{t-\tau }(\theta _{\tau }(Y)).  \label{eq:mult}
\end{equation}%
In formula \eqref{eq:mult}, $\theta _{\tau }$ is the shift operator that moves the time subscript
of $Y$ by $\tau $, that is, $(\theta _{\tau }(Y))_{s}= Y_{\tau + s}$.

We generalize  this construct by building an {\em extended multiplicative functional}.  The process $M$ is an extended multiplicative functional
if $M_0$ is a strictly positive (Borel measurable) function of $X_0$ and  $\{ { {M_t}/{M_0}}: t \in T \}$ is a multiplicative functional. This allows the process $M$ to be initialized at $M_0$ different from unity.
In this paper we drop the use of the term {\em extended} for pedagogical convenience.

%

\section{Perron--Frobenius Theory}\label{sec:app_PF}

%

%
\begin{proof}[\proofcaptionfont{Proof of Proposition}~\ref{prop:recover}]

Suppose there are two solutions $\widehat \eta, \widehat e$ and $\check \eta, \check e.$
Hence:
$$\exp(\widehat \eta t)\frac {\widehat H_t}{\widehat H_0}\frac{\widehat e (X_0)}{\widehat e(X_t)}=\exp(\check \eta t)\frac {\check H_t}{\check H_0}\frac{\check e(X_0)}{\check e (X_t)}$$
or if $k(x_t)=\frac{\check e (X_t)} {\widehat e (X_t)}>0,$ and $\eta=\check \eta-  \widehat \eta$
$$\exp(-\eta  t)\frac {\widehat H_t}{\widehat H_0}\frac{k(X_t)}{k(X_0)}= \frac {\check H_t}{\check H_0},$$
Calculating expected values in both sides and using the fact that  $\check H$ is a martingale we obtain that for $H=\widehat H$
\begin{equation}
E^H\left[ k(X_{t})\mid X_{0}=x\right] = \exp ( \eta t) k(x).  \label{eq:jose1}
\end{equation}
In what
follows we consider the discrete-time case. The continuous-time case uses an
identical approach, with the obvious changes.
First note that for a bounded function $f$ the Law of Large Numbers implies that
\begin{equation}\label{eq:lars1}
 \lim_{N\rightarrow \infty } \frac 1 {N}\sum_{t=1}^{N}E^H\left[ {f}(X_{t})\mid X_{0}=x\right] =E^H\left[ {f}(X_{0})%
\right] \end{equation}
for $H = \widehat H$ and $H=\check H.$
Consider three cases. First suppose that  $\eta<0$.
 Set
\begin{equation*}
{\widehat{k}}(x)=\min \left\{ 1,k(x)\right\} >0,\; \mbox{for all}\; x.
\end{equation*}%
Since  $\eta<0$, the right-hand side of \eqref{eq:jose1}  converges to zero for each $x$ as $%
t\rightarrow \infty. $ Thus, for $H=\widehat H$\begin{equation*}
0= \lim_{N\rightarrow \infty }{\frac{1}{N}}\sum_{t=1}^{N}E^H\left[
k(X_{t})\mid X_{0}=x\right] \geq \lim_{N\rightarrow \infty }{\frac{1}{N}}%
\sum_{t=1}^{N}E^H\left[ {\widehat{k}}(X_{t}) \mid X_{0}=x\right] =E^H\left[ {\widehat{k}}(X_{0})%
\right] >0.
\end{equation*}%
Thus we have established a contradiction.


Next suppose that  $\eta>0$. Note that for $H = \check H$
\begin{equation}
E^{H}\left[ {\frac{1}{k(X_{t})}}|X_{0}=x\right] = \exp ( -\eta t){\frac{1}{k(x)%
}}.  \label{eq:jose2}
\end{equation}%
Form
\begin{equation*}
{\widehat{k}}(x)=\min \left\{ 1,{\frac{1}{k(x)}}\right\}>0,\; \mbox{for all}\; x.
\end{equation*}%
Since  $\eta>0$, the right-hand side of \eqref{eq:jose2}  converges to zero for each $x$ as $%
t\rightarrow \infty. $ Thus,
\begin{equation*}
0= \lim_{N\rightarrow \infty }{\frac{1}{N}}\sum_{t=1}^{N}E^{H}\left[ {\frac{1}{%
k(X_{t})}}|X_{0}=x\right] \geq \lim_{N\rightarrow \infty }{\frac{1}{N}}%
\sum_{t=1}^{N}E^{H}\left[ {\widehat{k}}(X_{t})|X_{0}=x\right] =E^{H}\left[ {%
\widehat{k}}(X_{0})\right] >0.
\end{equation*}%
We have again established a contradiction.

Finally, suppose $\eta = 0$. Then again for for $H = \check H,$
\begin{equation*}
E^H\left[ \frac{1}{k(X_t)} \vert X_0 = x \right] = \frac{1}{k(x)}
\end{equation*}
for all $x.$   {From \eqref{eq:lars1}}:
\[
\lim_{N \rightarrow \infty} {\frac 1 N} \sum_{t=1}^{N} E^H\left[k^n(X_t) \vert X_0 = x \right] = E^H k^n(X_0).
\]
for $k^n = \min \{ 1/k, n \}$.  The same equality applies to the limit as $n\to\infty$ whereby $k^n$ is replaced by $1/k$.  Consequently,
$E^H\left[\frac{1}{k(X_0)}\right] = \frac{1}{k(x)}$ for almost all $x$.
  Thus $k(x)$ is a constant, and $\frac{\widehat H_t} {\widehat H_0} =\frac {\check H_t} {\check H_0}$ with probability one. \end{proof}

\section{Long-term valuation limits}\label{sec:app_valuation_limits}

First we verify that the approximation results  described in Section \ref{sec:longtermyield} hold under weaker conditions than Condition~\ref{stochstable2}. We assume instead:
\begin{equation} \label{liminf}
\liminf_{t \to \infty} {\widehat E} \left[ \psi \left(X_t\right) \mid X_0 = x \right]  > 0
\end{equation}
and that
\begin{equation}
\liminf_{t \to \infty} {\widehat E} \left[ \frac {\psi(X_t) }{\widehat e(X_t)}  \mid X_0 = x \right]  > 0.\label{liminf2}
\end{equation}
Notice that these assumptions follows if for every bounded function $f$
\begin{equation}
 \lim_{t\rightarrow \infty} {\widehat E} \left[ f(X_t) \mid X_0 = x \right] = {\widehat E}\left[ f(X_0) \right]\label{stochstabilityb}
\end{equation}
almost surely. \cite{meyntweedie} establish in Theorem 13.3.3 on page 327 that \eqref{stochstabilityb} holds for bounded functions $f$, provided $X$ is  aperiodic and positive Harris recurrent under the measure ${\widehat P}$.

Furthermore, we assume that
\[
 {\widehat E} \left[ {\frac { \psi(X_t) }{\widehat e(X_t)}}\right ]<\infty , \  \   {\widehat E} \left[ \psi(X_t) \right ] < + \infty,
\]
and that  $X$ satisfies Condition~\ref{cond:stationary} under ${\widehat P}$. Then,
\begin{eqnarray}\label{uglystuff}
{\frac 1 N} \log {\widehat E} \left[ \psi \left(X_N\right) \mid X_0 = x \right]   &\le &  {\frac 1 N}  \log \left(   {\widehat E}\left[  \sum_{t=1}^{N}
\psi \left(X_{t}\right) \vert X_0 = x \right] \right) \\ & = &
{\frac 1 N} \log N  + {\frac 1 N}  \log \left(   {\widehat E}\left[  {\frac 1 N} \sum_{t=1}^{N}
\psi\left(X_{t}\right) \vert X_0 = x \right] \right). \nonumber
\end{eqnarray}
Result \eqref{liminf}  implies that the $\liminf$ on the left-hand side converges to zero.  The $\limsup$ of the left-hand side also converges to zero.  To verify this, note that
the  Law of Large Numbers and the resulting almost sure convergence extend to the time-series averages of conditional expectations,
\[
\lim_{N \rightarrow \infty} {\widehat E}\left[  {\frac 1 N} \sum_{t=1}^{N}
\psi\left(X_{t}\right) \vert X_0 = x \right] = {\widehat E}\left [ \psi\left(X_{0}\right)\right]
\]
almost surely.
Hence both  terms on the right-hand side of \eqref{uglystuff} converge to zero with probability one.   Consequently,  $\limsup$ on the left-hand side does as well.  Given both the $\liminf$ and $\limsup$ on the left-hand side of \eqref{uglystuff} to zero the left-hand side must converge to zero almost surely.  

The same logic implies that
\[
\lim_{N \rightarrow \infty} {\frac 1 N} \log {\widehat E} \left[ {\frac {\psi(X_N)}{\widehat e(X_N)}} \mid X_0 = x \right] = 0,
\]
with probability one.\footnote{For the continuous-time case, sample the Markov process at integer points in time.  The process remains stationary under ${\widehat P}$, but is not necessarily ergodic.
The Law of Large Numbers still applies but with a limit that is the expectation conditioned on invariant events.   The previous argument with these modifications establishes the limiting behavior.}

Next introduce stochastic growth into the analysis as in Section \ref{sec:long-termrrt}. First note that the multiplicative functional $SG$ satisfies Condition~\ref{r:mrestrict}.  Let $ \eta^{\ast}$ denote the Perron--Frobenius eigenvalue using $SG$ in place of $S$ when solving Problem~\ref{PFP}.   This entails solving:
\[
 E\left[ S_t
 {G_t} {e^{\ast}}(X_t)  \mid X_0 = x \right] = \exp\left(  \eta^{\ast} t \right) {e^{\ast}}(x)
\]
and selecting  the eigenvalue-eigenfunction pair such that the implied martingale induces stationarity and ergodicity.  Imitating  our calculation of yields on securities with stationary payoffs
\[
E\left( S_t {G_t}  \left[{\frac {\widehat e(X_t)}{\widehat e(X_0)}} \right]  \mid X_0 = x \right) = \exp \left( t \eta^{\ast} \right) E^{{\ast}}    \left[{\frac {\widehat e(X_t)e^{\ast}(X_0) }{\widehat e(X_0) e^{\ast}(X_t) }}
\mid X_0 = x \right]
\]
where $E^{\ast}$ is constructed using the martingale $H^{{\ast}}$ with increments:
\[
{\frac {H_t^{{\ast}}}{H_0^{{\ast}}}} =  \exp\left( - \eta^{{\ast}} \right) S_t {G_t}\left[{\frac {e^{\ast}(X_t)}{e^{\ast}(X_0)}}\right].
\]
Similarly,
\[
E\left( S_t {G_t}    \mid X_0 = x \right) =  \exp \left( t \eta^{\ast} \right)  E^{{\ast}}  \left(   \left[{\frac {e^{\ast}(X_0) }{ e^{\ast}(X_t) }} \right] \mid X_0 = x \right).
\]
Suppose that $H^{{\ast}}$ induces a probability measure under which Condition~\ref{stochstable2} is satisfied.
In addition suppose that
\[
E^{\ast}\left[{\frac {\widehat e(X_t) }{ e^{\ast}(X_t) }} \right] < \infty,  \qquad   E^{\ast}\left[{\frac {1}{ e^{\ast}(X_t) }} \right] < \infty.
\]
From Condition~\ref{stochstable2} we obtain,
\begin{eqnarray*}
\lim_{t \rightarrow \infty} \frac1t E^{{\ast}}    \left[{\frac {\widehat e(X_t) }{ e^{\ast}(X_t) }}
\mid X_0 = x \right] &= & 0 \\
\lim_{t \rightarrow \infty} \frac1t  \log E^{{\ast}}    \left[{\frac {1 }{ e^{{\ast}}(X_t) }}
\mid X_0 = x \right]  &= & 0.
\end{eqnarray*}
Thus
\begin{eqnarray*}
&&\lim_{t \rightarrow \infty} \frac1t \log  E\left( S_t {G_t}  \left[{\frac {\widehat e(X_t)}{\widehat e(X_0)}} \right]  \mid X_0 = x \right) = \\
&&\qquad = \eta^{\ast} +  \lim_{t \rightarrow \infty}{\frac 1 t} \left[  \log e^{\ast}(x) ) - \log \widehat e(x) \right] + \lim_{t \rightarrow \infty} \frac1t E^{{\ast}}    \left[{\frac {\widehat e(X_t) }{ e^{\ast}(X_t) }}
\mid X_0 = x \right]
= \eta^{\ast} ,
\end{eqnarray*}
and similarly
\begin{eqnarray*}
&&\lim_{t \rightarrow \infty} \frac1t \log  E\left( S_t {G_t}    \mid X_0 = x \right)= \\
&&\qquad = \eta^{\ast} + \lim_{t \rightarrow \infty} {\frac 1 t} \log e^{{\ast}}(x)  +
\lim_{t \rightarrow \infty} \frac1t  \log E^{{\ast}}    \left[{\frac {1 }{ e^{{\ast}}(X_t) }}
\mid X_0 = x \right] = \eta^{\ast}.
\end{eqnarray*}
Hence
\[
\lim_{t\rightarrow \infty} y_t[G](x) = - \eta + \eta^{\ast} - \eta^{\ast} = - \eta.
\]

\section{Derivations for the model with predictable consumption dynamics}\label{sec:app_LRR_details}


In this section, we provide derivations for the model analyzed in Section~\ref{sec:quantitative}. We will focus on the analysis of the Perron--Frobenius problem. A complete analysis of the model can be found in \cite{hansen:2012} and in the appendix in \cite{borovicka_hansen_scheinkman:2014}.

\subsection{Martingale decomposition}\label{sec:app_martingale_decomposition}

We solve the Perron--Frobenius problem
\begin{equation*}
E\left[ S_{t}\widehat{e}\left( X_{t}\right) \mid X_{0}=x\right] =\exp \left( \widehat\eta
t\right) \widehat{e}\left( x\right)
\end{equation*}%
where $S$ is a multiplicative functional parameterized by coefficients $(\beta_s(x),\alpha_s(x))$. Since the problem holds for every $t$, then it also holds in the limit (as long as it exists)
\begin{equation*}
\lim_{t\searrow0}\frac{1}{t}\left[ E\left[ S_{t}\widehat{e}\left( X_{t}\right) \mid X_{0}=x\right] -\exp \left( \widehat\eta
t\right) \widehat{e}\left( x\right) \right] = 0.
\end{equation*}
The limit yields the partial differential equation
\begin{equation}
\mathbb{S}e=\eta e  \label{eq:generator_equation}
\end{equation}%
where, in the general Brownian information setup given by equations (\ref{eq:dlogM_ct}), the infinitesimal generator $\mathbb{%
S}$ is given by%
\begin{equation*}
\mathbb{S}\widehat{e}=\left( \beta_s +\frac{1}{2}\left\vert \alpha_s \right\vert
^{2}\right) \widehat{e}+\widehat{e}_{x}\cdot \left( \mu +\sigma \alpha_s \right) +\frac{1}{2}%
\mathrm{tr}\left[ \widehat{e}_{xx}\sigma \sigma ^{\prime }\right] .
\end{equation*}
Equation (\ref{eq:generator_equation}) is therefore a second-order partial differential
equation, and we are looking for a solution in the form of a number $\widehat\eta $
and a strictly positive function $\widehat{e}$. \cite{hansen_scheinkman:2009} show
that if there are multiple such solutions, then only the solution associated
with the lowest value of $\widehat\eta $ can generate ergodic dynamics
under the implied change of measure.

In the case of the long-run risk model introduced in Section~\ref%
{sec:quantitative} and parameterized by (\ref{eq:LRR_X_params})--(\ref{eq:LRR_M_params}), we can guess%
\begin{equation*}
\widehat{e}\left( x\right) =\exp\left( \bar{e}_{1} x_{1}+\bar{e}_{2}x_{2}\right)
\end{equation*}%
which leads to the system of equations%
\begin{eqnarray*}
\eta  &=&{\bar{\beta}}_{s,0}-{\bar{\beta}}_{s,11} \iota _{1}-{\bar{\beta}}%
_{s,12}\iota _{2}-\bar{e}_{1} \left( {\bar{\mu}}_{11}\iota _{1}+{\bar{\mu}%
}_{12}\iota _{2}\right) -\bar{e}_{2}{\bar{\mu}}_{22}\iota _{2} \\
0 &=&{\bar{\beta}}_{s,11}+\bar{\mu}_{11}\bar{e}_{1} \\
0 &=&{\bar{\beta}}_{s,12}+\frac{1}{2}\left\vert \bar{\alpha}_s\right\vert ^{2}+%
\bar{e}_{1} \left( {\bar{\mu}}_{12}+\bar{\sigma}_{1}\bar{\alpha}_s\right)
+\frac{1}{2}\bar{e}_{1}^{2 }\left\vert\bar{\sigma}_{1}\right\vert^2%
+
\bar{e}_{2}\left( {\bar{\mu}}_{22}+\bar{\sigma}_{2}\bar{\alpha}_s+\bar{e}%
_{1}\bar{\sigma}_{1}\bar{\sigma}_{2}^{\prime }\right) +\frac{1}{2}%
\left( \bar{e}_{2}\right) ^{2}\left\vert\bar{\sigma}_{2}\right\vert^2.
\end{eqnarray*}%
The coefficients $\bar\beta_{s,\cdot}$ and $\bar\alpha_s$ will be determined below. The last equation is a quadratic equation for $\bar{e}_{2}$ and we choose
the solution for $\bar{e}_{2}$ that leads to the smaller value of $\widehat\eta $.\
From the decomposition%
\begin{equation*}
S_{t}=\exp \left( \widehat\eta t\right) \frac{\widehat{e}\left( X_{0}\right)
}{\widehat{e}\left( X_{t}\right) } \frac{\widehat{H}_{t}}{\widehat{H}_{0}}
\end{equation*}%
we can extract the martingale $\widehat{H}$:%
\begin{equation*}
d\log \widehat{H}_{t}=d\log S_{t}-\widehat\eta dt+d\log \widehat{e}\left( X_{t}\right)
\end{equation*}%
and thus%
\begin{equation*}
\frac{d\widehat{H}_{t}}{\widehat{H}_{t}}=\sqrt{X_{2t}}\left( \bar{\alpha}%
_{s}+\bar{\sigma}_{1}^{\prime }\bar{e}_{1}+\bar{\sigma}_{2}^{\prime }\bar{e}%
_{2}\right) \cdot dW_{t}\dot{=}\sqrt{X_{2t}}\widehat{\alpha }_{h}\cdot
dW_{t}.
\end{equation*}%
This martingale implies a change of measure such that $\widehat{W}$
defined as%
\begin{equation*}
d\widehat{W}_{t}=-\sqrt{X_{2t}}\widehat{\alpha }_{h}dt+dW_{t}
\end{equation*}%
is a Brownian motion under the new measure $\widehat{P}$. Under the change
of measure implied by $\widehat{H}$, we can write the joint dynamics of
the model as%
\begin{eqnarray*}
dX_{1t} &=&\left[ \widehat{\mu }_{11}\left( X_{1t}-\widehat{\iota }%
_{1}\right) +\widehat{\mu }_{12}\left( X_{2t}-\widehat{\iota }%
_{2}\right) \right] dt+\sqrt{X_{2t}}\bar{\sigma}_{1}d\widehat{W}_{t} \\
dX_{2t} &=&\widehat{\mu }_{22}\left( X_{2t}-\widehat{\iota }_{2}\right)
dt+\sqrt{X_{2t}}\bar{\sigma}_{2}d\widehat{W}_{t}
\end{eqnarray*}%
where%
\begin{eqnarray*}
\widehat{\mu }_{11} &=&\bar{\mu}_{11}\qquad \widehat{\mu }_{12}=\bar{\mu}%
_{12}+\bar{\sigma}_{1}\widehat{\alpha }_h\qquad \widehat{\mu }_{22}=\bar{%
\mu}_{22}+\bar{\sigma}_{2}\widehat{\alpha }_h \\
\widehat{\iota }_{2} &=&\frac{\bar{\mu}_{22}}{\widehat{\mu }_{22}}\iota
_{2}\qquad \widehat{\iota }_{1}=\iota _{1}+\left( \bar{\mu}_{11}\right)
^{-1}\left( \bar{\mu}_{12}\iota _{2}-\widehat{\mu }_{12}\widehat{\iota }%
_{2}\right) .
\end{eqnarray*}

Similarly, every multiplicative functional $M$ with parameters given by (\ref%
{eq:LRR_M_params}) can be rewritten as%
\begin{equation*}
d\log M_{t}=\left[ \widehat{\beta }_{0}+\widehat{\beta }_{11}
\left( X_{1t}-\widehat{\iota }_{1}\right) +\widehat{\beta }_{12}\left(
X_{2t}-\widehat{\iota }_{2}\right) \right] dt+\sqrt{X_{2t}}\bar{\alpha}%
\cdot d\widehat{W}_{t}
\end{equation*}%
where%
\begin{eqnarray*}
\widehat{\beta }_{0} &=&\bar{\beta}_{0}+\bar{\beta}_{11} \left(
\widehat{\iota }_{1}-\iota _{1}\right) +\bar{\beta}_{12}\left( \widehat{%
\iota }_{2}-\iota _{2}\right) +\left( \bar{\alpha}\cdot \widehat{\alpha }_h%
\right) \widehat{\iota }_{2} \\
\widehat{\beta }_{11} &=&\bar{\beta}_{11}\qquad \widehat{\beta }_{12}=%
\bar{\beta}_{12}+\bar{\alpha}\cdot \widehat{\alpha }_h\text{.}
\end{eqnarray*}


\subsection{Value function and stochastic discount factor for recursive
utility}

\label{sec:app_lrr_SDF}We choose a convenient choice for representing
continuous values. Similar to the discussion in \cite{schroder_skiadas:1999}%
, we use the counterpart to discounted expected logarithmic utility.
\begin{equation*}
d\log  V_{t}=\mu _{v,t}dt+\sigma _{v,t}\cdot dW_{t}.
\end{equation*}%
The local evolution satisfies:
\begin{equation}
\mu _{v,t}=\delta \log V_{t}-\delta \log C_{t}-{\frac{1-\gamma }{2}}|\sigma
_{v,t}|^{2}  \label{appen1}
\end{equation}%
When $\gamma =1$ this collapses to the discounted expected utility recursion.

Let
\begin{equation*}
\log V_t = \log C_t + v(X_t)
\end{equation*}
and guess that
\begin{equation*}
v\left( x\right) =\bar{v}_{0}+\bar{v}_{1}\cdot x_{1}+\bar{v}_{2}x_{2}.
\end{equation*}
We  compute $\mu_{v,t}$ by applying the infinitesimal generator to $\log
C + v(X)$. In addition,
\begin{equation*}
\sigma_{v,t} = \alpha _{c} \left( X_t\right) +\sigma \left( X_t \right)
^{\prime }\frac{\partial }{\partial x}v\left( X_t \right).
\end{equation*}
Substituting into \eqref{appen1} leads to a set of algebraic equations%
\begin{eqnarray*}
\delta \bar{v}_{0} &=&\bar{\beta}_{c,0}-\iota
_{1}\left( \bar{\beta}_{c,1}+\bar{\mu}_{11}\bar{v}_{1}\right)
-\iota _{2}\left( \bar{\beta}_{c,2}+\bar{\mu}_{12}\bar{v}_{1}+\bar{\mu}_{22}%
\bar{v}_{2}\right)  \\
\delta \bar{v}_{1} &=&\bar{\beta}_{c,1}+\bar{\mu}_{11}\bar{v}_{1}
\\
\delta \bar{v}_{2} &=&\bar{\beta}_{c,2}+\bar{\mu}_{12}\bar{v}_{1}+%
\bar{\mu}_{22}\bar{v}_{2}+\frac{1}{2}\left( 1-\gamma \right) \left\vert \bar{%
\alpha}_{c}+\bar{\sigma}_{1}^{\prime }\bar{v}_{1}+\bar{\sigma}_{2}^{\prime }%
\bar{v}_{2}\right\vert ^{2}
\end{eqnarray*}%
which can be solved for the coefficients $\bar{v}_{i}$. The third equation
is a quadratic equation for $\bar{v}_{2}$ that has a real solution if and
only if%
\begin{eqnarray*}
D &=&\left[ \bar{\mu}_{22}-\delta +\left( 1-\gamma \right) \left( \bar{\alpha%
}_{c}+\bar{\sigma}_{1}^{\prime }\bar{v}_{1}\right) ^{\prime }\bar{\sigma}%
_{2}^{\prime }\right] ^{2}- \\
&&-2\left( 1-\gamma \right) \left\vert \bar{\sigma}_{2}\right\vert
^{2}\left( \bar\beta _{c,2}+\bar\mu _{12}\bar{v}_{1}+\frac{1}{2}\left(
1-\gamma \right) \left\vert \bar{\alpha}_{c}+\bar{\sigma}_{1}^{\prime }\bar{v%
}_{1}\right\vert ^{2}\right) \geq 0.
\end{eqnarray*}%
In particular, the solution will typically not exist for large values of $%
\gamma $. If the solution exists, it is given by%
\begin{equation}
\bar{v}_{2}=\frac{-\left[ \bar{\mu}_{22}-\delta +\left( 1-\gamma \right)
\left( \bar{\alpha}_{c}+\bar{\sigma}_{1}^{\prime }\bar{v}_{1}\right)
^{\prime }\bar{\sigma}_{2}^{\prime }\right] \pm \sqrt{D}}{\left( 1-\gamma
\right) \left\vert \bar{\sigma}_{2}^{\prime }\right\vert ^{2}}
\label{eq:v2bar}
\end{equation}%
The solution with the minus sign is the one that interests us.

The resulting stochastic discount factor has two components. One component is the
intertemporal marginal rate of substitution for discounted log utility and
the other is a martingale constructed from the continuation value
\begin{equation*}
d\log S_{t}=- \delta dt -d\log C_{t}+d\log H^*_{t}
\end{equation*}%
where $H^*$ is a martingale given by%
\begin{equation*}
\frac{d H^*_{t}}{H^*_{t}}=\sqrt{X_{2,t}}\left( 1-\gamma
\right) \left( \bar{\alpha}_{c}+\bar{\sigma}_{1}^{\prime }\bar{v}_{1}+\bar{%
\sigma}_{2}^{\prime }\bar{v}_{2}\right) ^{\prime }dW_{t}.
\end{equation*}%
This determines the coefficients $(\beta_s(x), \alpha_s(x))$ of the stochastic discount factor.

When we choose the `minus' solution in equation (\ref{eq:v2bar}), then $%
\widehat{S}$ implies a change of measure that preserves ergodicity. Notice that while $H^*$ is a martingale, it is distinct from $\widehat{H}_t / \widehat{H}_0$ as long as the consumption process itself contains a nontrivial martingale component.

\subsection{Conditional expectations of multiplicative functionals}

In order to compute asset prices and their expected returns, we need to compute conditional expectations of multiplicative functionals $M$ parameterized by (\ref{eq:LRR_X_params})--(\ref{eq:LRR_M_params}). These conditional expectations are given by
\begin{equation*}
E\left[ M_{t}\mid X_{0}=x\right] =\exp \left[ \theta _{0}(t)+\theta
_{1}(t)\cdot x_{1}+\theta _{2}(t)x_{2}\right] \text{,}
\end{equation*}%
where parameters $\theta_i(t)$ satisfy a system of ordinary differential equations derived in \cite{hansen:2012} and in the appendix of \cite{borovicka_hansen_scheinkman:2014}.

\ifthenelse{\equal{\compiletoJFstyle}{1}}{}{%
  \sectionfont{\Large}
  \subsectionfont{\large}
  \end{small}
}

\clearpage

\ifthenelse{\equal{\compiletoJFstyle}{1}}{
  \begin{doublespacing}   
  \bibliographystyle{jf}
  \bibliography{recoverybib}
  \end{doublespacing}
}{%
  \bibliography{recoverybib}

\begin{thebibliography}{49}
\providecommand{\natexlab}[1]{#1}
\providecommand{\selectlanguage}[1]{\relax}

\bibitem[{A{\"i}t-Sahalia and Lo(2000)}]{AitsahaliaLo2000}
A{\"i}t-Sahalia, Yacine and Andrew Lo. 2000.
\newblock Nonparametric risk management and implied risk aversion.
\newblock \emph{Journal of Econometrics} 94~(1-2):9--51.

\bibitem[{Almeida and Garcia(2013)}]{almeidagarcia13}
Almeida, Caio and Ren{\'{e}} Garcia. 2013.
\newblock Robust Economic Implications of Nonlinear Pricing Kernels.

\bibitem[{Alvarez and Jermann(2005)}]{alvarez_jermann:2005}
Alvarez, Fernando and Urban~J. Jermann. 2005.
\newblock Using Asset Prices to Measure the Persistence of the Marginal Utility
  of Wealth.
\newblock \emph{Econometrica} 73~(6):1977--2016.

\bibitem[{Anderson et~al.(2003)Anderson, Hansen, and
  Sargent}]{anderson_hansen_sargent:2003}
Anderson, Evan~W., Lars~Peter Hansen, and Thomas~J. Sargent. 2003.
\newblock A Quartet of Semigroups for Model Specification, Robustness, Prices
  of Risk, and Model Detection.
\newblock \emph{Journal of the European Economic Association} 1~(1):68--123.

\bibitem[{Backus et~al.(1989)Backus, Gregory, and Zin}]{bgz}
Backus, David~K., Allan~W. Gregory, and Stanley~E. Zin. 1989.
\newblock Risk Premiums in the Term Structure : Evidence from Artificial
  Economies.
\newblock \emph{Journal of Monetary Economics} 24~(3):371--399.

\bibitem[{Backus et~al.(2014)Backus, Chernov, and
  Zin}]{backus_chernov_zin:2014}
Backus, David~K., Mikhail Chernov, and Stanley~E. Zin. 2014.
\newblock Sources of Entropy in Representative Agent Models.
\newblock \emph{The Journal of Finance} 59~(1):51--99.

\bibitem[{Bakshi and Chabi-Yo(2012)}]{bakshi_chabiyo:2012}
Bakshi, Gurdip and Fousseni Chabi-Yo. 2012.
\newblock Variance Bounds on the Permanent and Transitory Components of
  Stochastic Discount Factors.
\newblock \emph{Journal of Financial Economics} 105~(1):191--208.

\bibitem[{Bansal and Lehmann(1994)}]{bansal_lehmann:1994}
Bansal, Ravi and Bruce~N. Lehmann. 1994.
\newblock Addendum to `Growth-Optimal Portfolio Restrictions on Asset Pricing
  Models'.
\newblock Unpublished manuscript.

\bibitem[{Bansal and Lehmann(1997)}]{bansallehmann}
---{}---{}---. 1997.
\newblock Growth-Optimal Portfolio Restrictions on Asset Pricing Models.
\newblock \emph{Macroeconomic Dynamics} 1~(2):333--354.

\bibitem[{Bansal and Yaron(2004)}]{bansal_yaron:2004}
Bansal, Ravi and Amir Yaron. 2004.
\newblock Risks for the Long Run: A Potential Resolution of Asset Pricing
  Puzzles.
\newblock \emph{Journal of Finance} 59~(4):1481--1509.

\bibitem[{Bidder and Smith(2013)}]{bidder_smith:2013}
Bidder, Rhys and Matthew~E. Smith. 2013.
\newblock Doubts and Variability: {A} Robust Perspective on Exotic Consumption
  Series.
\newblock Federal Reserve Bank of San Francisco Working Paper 2013--28.

\bibitem[{Borovi\v{c}ka et~al.(2011)Borovi\v{c}ka, Hansen, Hendricks, and
  Scheinkman}]{bhhs11}
Borovi\v{c}ka, Jaroslav, Lars~Peter Hansen, Mark Hendricks, and Jos\'{e}~A.
  Scheinkman. 2011.
\newblock Risk Price Dynamics.
\newblock \emph{Journal of Financial Econometrics} 9~(1):3--65.

\bibitem[{Borovi\v{c}ka et~al.(2014)Borovi\v{c}ka, Hansen, and
  Scheinkman}]{borovicka_hansen_scheinkman:2014}
Borovi\v{c}ka, Jaroslav, Lars~Peter Hansen, and Jos\'{e}~A. Scheinkman. 2014.
\newblock Shock Elasticities and Impulse Responses.
\newblock \emph{Mathematics and Financial Economics} 8~(4):333--354.

\bibitem[{Breeden and Litzenberger(1978)}]{breedenlitzenberger78}
Breeden, Douglas~T. and Robert~H. Litzenberger. 1978.
\newblock Prices of State-Contingent Claims Implicit in Option Prices.
\newblock \emph{The Journal of Business} 51~(4):621--51.

\bibitem[{Breiman(1982)}]{breiman}
Breiman, Leo. 1982.
\newblock \emph{Probability. {\rm Volume 7 of Classics in Applied
  Mathematics}}.
\newblock Philadelphia, PA: Society for Industrial and Applied Mathematics
  (SIAM).

\bibitem[{Carr and Yu(2012)}]{carr_yu:2012}
Carr, Peter and Jiming Yu. 2012.
\newblock Risk, Return and {R}oss Recovery.
\newblock \emph{The Journal of Derivatives} 20~(1):38--59.

\bibitem[{Chen et~al.(2014)Chen, Chernozhukov, Lee, and Newey}]{CCLN:2014}
Chen, Xiaohong, Victor Chernozhukov, Sokbae Lee, and Whitney~K. Newey. 2014.
\newblock Local Identification of Nonparametric and Semiparametric Models.
\newblock \emph{Econometrica} 82~(2):785--809.

\bibitem[{Christensen(2014)}]{christiansen:14}
Christensen, Timothy~M. 2014.
\newblock Nonparametric Indentification of Positive Eigenfunctions.
\newblock \emph{Econometric Theory} FirstView:1--21.

\bibitem[{Cressie and Read(1984)}]{cressieread84}
Cressie, Noel and Timothy R.~C. Read. 1984.
\newblock Multinomial Goodness-of-Fit Tests.
\newblock \emph{Journal of the Royal Statistical Society. Series B
  (Methodological)} 46~(3):440--464.

\bibitem[{Dubynskiy and Goldstein(2013)}]{dubynskiy_goldstein:2013}
Dubynskiy, Sergey and Robert~S. Goldstein. 2013.
\newblock Recovering Drifts and Preference Parameters from Financial
  Derivatives.

\bibitem[{Duffie and Epstein(1992)}]{duffie_epstein:1992}
Duffie, Darrell and Larry~G. Epstein. 1992.
\newblock Asset Pricing with Stochastic Differential Utility.
\newblock \emph{The Review of Financial Studies} 5~(3):411--436.

\bibitem[{Epstein and Zin(1989)}]{epstein_zin:1989}
Epstein, Larry~G. and Stanley~E. Zin. 1989.
\newblock Substitution, Risk Aversion, and the Temporal Behavior of Consumption
  and Asset Returns: A Theoretical Framework.
\newblock \emph{Econometrica} 57~(4):937--969.

\bibitem[{Gagliardini et~al.(2011)Gagliardini, Gourieroux, and
  Renault}]{GGR:2011}
Gagliardini, Patrick, Christian Gourieroux, and Eric Renault. 2011.
\newblock Efficient Derivative Pricing by the Extended Method of Moments.
\newblock \emph{Econometrica} 79~(4):1181--1232.

\bibitem[{Garcia et~al.(2010)Garcia, Ghysels, and Renault}]{GGR2010}
Garcia, Ren{\'e}, Eric Ghysels, and Eric Renault. 2010.
\newblock The Econometrics of Option Pricing.
\newblock In \emph{Handbook of Financial Econometrics: Tools and Techniques},
  vol.~1 of \emph{Handbooks in Finance}, edited by Lars~Peter Hansen and Yacine
  A{\"i}t-Sahalia, chap.~9, 479 -- 552. San Diego: North-Holland.

\bibitem[{Ghosh et~al.(2012)Ghosh, Julliard, and Taylor}]{gjt12}
Ghosh, Anisha, Christian Julliard, and Alex Taylor. 2012.
\newblock What is the Consumption-CAPM Missing? An Information-Theoretic
  Framework for the Analysis of Asset Pricing Models.
\newblock Working paper, Tepper School of Business, Carnegie Mellon University.

\bibitem[{Hansen(2012)}]{hansen:2012}
Hansen, Lars~Peter. 2012.
\newblock Dynamic Valuation Decomposition within Stochastic E\-co\-no\-mies.
\newblock \emph{Econometrica} 80~(3):911--967.
\newblock Fisher--Schultz Lecture at the European Meetings of the Econometric
  Society.

\bibitem[{Hansen(2014)}]{hansennobel:14}
---{}---{}---. 2014.
\newblock Nobel Lecture: {U}ncertainty Outside and Inside Economic Models.
\newblock \emph{Journal of Political Economy} 122~(5):945--987.

\bibitem[{Hansen and Jagannathan(1991)}]{hansenjagannathan91}
Hansen, Lars~Peter and Ravi Jagannathan. 1991.
\newblock Implications of Security Market Data for Models of Dynamic Economies.
\newblock \emph{Journal of Political Economy} 99~(2):225--262.

\bibitem[{Hansen and Richard(1987)}]{hansen_richard:1987}
Hansen, Lars~Peter and Scott~F. Richard. 1987.
\newblock The Role of Conditioning Information in Deducing Testable
  Restrictions Implied by Dynamic Asset Pricing Models.
\newblock \emph{Econometrica} 55~(3):587--613.

\bibitem[{Hansen and Scheinkman(2009)}]{hansen_scheinkman:2009}
Hansen, Lars~Peter and Jos\'{e}~A. Scheinkman. 2009.
\newblock Long Term Risk: An Operator Approach.
\newblock \emph{Econometrica} 77~(1):177--234.

\bibitem[{Hansen and Scheinkman(2014)}]{hansen_scheinkman:2014}
---{}---{}---. 2014.
\newblock Stochastic Compounding and Uncertain Valuation.

\bibitem[{Hansen and Singleton(1982)}]{hansensingleton82}
Hansen, Lars~Peter and Kenneth~J. Singleton. 1982.
\newblock Generalized Instrumental Variables Estimation of Nonlinear Rational
  Expectations Models.
\newblock \emph{Econometrica} 50~(5):1269--1286.

\bibitem[{Hansen et~al.(1995)Hansen, Heaton, and Luttmer}]{hhl95}
Hansen, Lars~Peter, John Heaton, and Erzo G.~J. Luttmer. 1995.
\newblock Econometric Evaluation of Asset Pricing Models.
\newblock \emph{The Review of Financial Studies} 8~(2):237--274.

\bibitem[{Hansen et~al.(2007)Hansen, Heaton, Lee, and Roussanov}]{hhlr}
Hansen, Lars~Peter, John~C. Heaton, Junghoon Lee, and Nikolai Roussanov. 2007.
\newblock Intertemporal Substitution and Risk Aversion.
\newblock In \emph{Handbook of Econometrics, \rm vol 6A}, edited by James~J.
  Heckman and Edward~E. Leamer, 3967--4056. Amsterdam: North-Holland.

\bibitem[{Hansen et~al.(2008)Hansen, Heaton, and Li}]{hhl}
Hansen, Lars~Peter, John~C. Heaton, and Nan Li. 2008.
\newblock Consumption Strikes Back? Measuring Long-Run Risk.
\newblock \emph{Journal of Political Economy} 116:260--302.

\bibitem[{Harrison and Kreps(1979)}]{harrisonkreps79}
Harrison, J.~Michael and David~M. Kreps. 1979.
\newblock {Martingales and Arbitrage in Multiperiod Securities Markets}.
\newblock \emph{Journal of Economic Theory} 20~(3):381--408.

\bibitem[{Hilscher et~al.(2014)Hilscher, Raviv, and Reis}]{hrr:2014}
Hilscher, Jens, Alon Raviv, and Ricardo Reis. 2014.
\newblock Inflating Away the Public Debt?
\newblock Columbia University.

\bibitem[{Kazemi(1992)}]{kazemi:1992}
Kazemi, Hossein~B. 1992.
\newblock An Intertemporal Model of Asset Prices in a {M}arkov Economy with a
  Limiting Stationary Distribution.
\newblock \emph{Review of Financial Studies} 5~(1):85--104.

\bibitem[{Kreps and Porteus(1978)}]{kreps_porteus:1978}
Kreps, David~M. and Evan~L. Porteus. 1978.
\newblock Temporal Resolution of Uncertainty and Dynamic Choice Theory.
\newblock \emph{Econometrica} 46~(1):185--200.

\bibitem[{Meyn and Tweedie(1993)}]{meyntweedieII}
Meyn, Sean~P. and Richard~L. Tweedie. 1993.
\newblock Stability of {M}arkovian Processes {II}: Continuous-Time Processes
  and Sampled Chains.
\newblock \emph{Advances in Applied Probability} 25~(3):487--517.

\bibitem[{Meyn and Tweedie(2009)}]{meyntweedie}
---{}---{}---. 2009.
\newblock \emph{Markov Chains and Stochastic Stability}.
\newblock Cambridge, United Kingdom: Cambridge University Press.

\bibitem[{Park(2014)}]{park:2014}
Park, Hyungbin. 2014.
\newblock Ross Recovery with Recurrent and Transient Processes.

\bibitem[{Qin and Linetsky(2014{\natexlab{a}})}]{qin_linetsky:2014_1}
Qin, Likuan and Vadim Linetsky. 2014{\natexlab{a}}.
\newblock Long Term Risk: {A} Martingale Approach.
\newblock Mimeo, Northwestern University.

\bibitem[{Qin and Linetsky(2014{\natexlab{b}})}]{qin_linetsky:2014_2}
---{}---{}---. 2014{\natexlab{b}}.
\newblock Positive Eigenfunctions of {M}arkovian Pricing Operators:
  {H}an{\-}sen--{S}cheink{\-}man Factorization and {R}oss Recovery.
\newblock Mimeo, Northwestern University.

\bibitem[{Ross(1978)}]{ross78}
Ross, Stephen~A. 1978.
\newblock {A Simple Approach to the Valuation of Risky Streams}.
\newblock \emph{The Journal of Business} 51~(3):453--75.

\bibitem[{Ross(2015)}]{ross:2015}
Ross, Stephen~A. 2015.
\newblock The Recovery Theorem.
\newblock \emph{Journal of Finance} 70~(2):615--648.

\bibitem[{Schroder and Skiadas(1999)}]{schroder_skiadas:1999}
Schroder, Mark and Costis Skiadas. 1999.
\newblock Optimal Consumption and Portfolio Selection with Stochastic
  Differential Utility.
\newblock \emph{Journal of Economic Theory} 89~(1):68--126.

\bibitem[{Stutzer(1996)}]{stutzer96}
Stutzer, Michael. 1996.
\newblock A Simple Nonparametric Approach to Derivative Security Valuation.
\newblock \emph{The Journal of Finance} 51~(5):1633--1652.

\bibitem[{Walden(2014)}]{walden:2014}
Walden, Johan. 2014.
\newblock Recovery with Unbounded Diffusion Processes.

\end{thebibliography}
}
\ifthenelse{\equal{\compiletoJFstyle}{1}}{
  \clearpage

  \renewcommand{\enotesize}{\normalsize}
  \begin{doublespacing}
    \theendnotes
  \end{doublespacing}
}{}
\ifthenelse{\equal{\compiletoJFstyle}{1}}{
  \clearpage
  \begin{figure}[!t]
\begin{center}
\myPlotXDistUndistorted\myPlotXDistDistorted
\end{center}
\vspace{-4mm}
\caption{{\bf Stationary densities for the state vector $X=(X_1,X_2)^\prime$ under the correctly specified probability measure $P$ (left panel) and the recovered probability measure $\widehat{P}$ (right panel).} The dashed line in the right panel corresponds to the outermost contour for the distribution under the risk neutral probability measure $\overline{P}$.
The parameterization of the model
is $\bar{\protect\beta}_{c,0}=0.0015$, $\bar{\protect\beta}_{c,1}=1$, $\bar{%
\protect\beta}_{c,2}=0$, $\bar{\protect\mu}_{11}=-0.021$, $\bar{\protect\mu}%
_{12}=\bar{\protect\mu}_{21}=0$, $\bar{\protect\mu}_{22}=-0.013$, $\bar{%
\protect\alpha}_{c}=\left[ 0.0078~~0~~0\right] ^{\prime }$, $\bar{\protect%
\sigma}_{1}=\left[ 0~~0.00034~~0\right] $, $\bar{\protect\sigma}_{2}=\left[
0~~0~~-0.038\right] $, $\protect\iota _{1}=0$, $\protect\iota _{2}=1$, $%
\protect\delta =0.002$, $\protect\gamma =10$. Parameters are calibrated to
monthly frequency.}
\label{fig:X_distribution_latex}
\end{figure}

  \begin{figure}[!t]
\begin{center}
\myPlotYields\myPlotBondYields
\end{center}
\vspace{-4mm}
\ifthenelse{\equal{\compiletoJFstyle}{1}}{
\caption{{\bf Yields under the true and recovered probability measure.} The graphs show the annualized yields on cash flows corresponding to the aggregate consumption process (left panel) and on bonds (right panel) with different maturities. The dark bands with solid lines correspond to the distribution under~$P$, while the light bands with dashed lines to the distribution under~$\widehat P$. The lines represent quartiles of the distribution. The parameterization is as in Figure~\ref{fig:X_distribution_latex}.}
}{
\caption{{\bf Yields under the true and recovered probability measure.} The graphs show the annualized yields on cash flows corresponding to the aggregate consumption process (left panel) and on bonds (right panel) with different maturities. The blue bands with solid lines correspond to the distribution under~$P$, while the red bands with dashed lines to the distribution under~$\widehat P$. The lines represent quartiles of the distribution. The parameterization is as in Figure~\ref{fig:X_distribution_latex}.}
}
\label{fig:LRR_asset_pricing}
\end{figure}


}{}

\end{document}